\documentclass[twocolumn]{aastex61}  % USE THIS TO MAKE BIB, THEN FORMAT USING EMULATEAPJ

\usepackage{mathtools}
\usepackage{amsmath}
\usepackage{graphicx}
\usepackage{natbib}
\usepackage[all]{hypcap}								% Workaround for hyperref
\usepackage{hyperref}
\usepackage{xcolor}
\usepackage{braket}
\usepackage{fancyhdr}
\usepackage{bm}
\hypersetup{linkcolor=cyan,colorlinks=true,filecolor=cyan,citecolor=cyan}

\def\tocm{$21\,\textrm{cm}$\ }

\def\kpara{k_{\parallel}}
\def\kperp{k_{\perp}}

\begin{document}

\title{Absolute Calibration Strategies for the Hydrogen Epoch of Reionization Array and Their Impact on the 21\,cm Power Spectrum}
\shorttitle{Absolute Calibration Strategies for HERA}
\shortauthors{Kern et al.}

% Corresponding Author
\correspondingauthor{Nicholas Kern}
\email{nkern@berkeley.edu}
\author{Nicholas S. Kern}
\affiliation{Department of Astronomy, University of California, Berkeley, CA}

% Contributing Authors
\author{Joshua S. Dillon}
\affiliation{Department of Astronomy, University of California, Berkeley, CA}
\affiliation{NSF Astronomy and Astrophysics Postdoctoral Fellow}

\author{Aaron R. Parsons}
\affiliation{Department of Astronomy, University of California, Berkeley, CA}

\author{Christopher L. Carilli}
\affiliation{National Radio Astronomy Observatory, Socorro, NM}

\author{Gianni  Bernardi}
\affiliation{Department of Physics and Electronics, Rhodes University, PO Box 94, Grahamstown, 6140, South Africa}
\affiliation{INAF-Istituto di Radioastronomia, via Gobetti 101, 40129 Bologna, Italy}
\affiliation{SKA-SA, Cape Town, South Africa}

\author{Zara  Abdurashidova}
\affiliation{Department of Astronomy, University of California, Berkeley, CA}

\author{James E. Aguirre}
\affiliation{Department of Physics and Astronomy, University of Pennsylvania, Philadelphia, PA}

\author{Paul  Alexander}
\affiliation{Cavendish Astrophysics, University of Cambridge, Cambridge, UK}

\author{Zaki S. Ali}
\affiliation{Department of Astronomy, University of California, Berkeley, CA}

\author{Yanga  Balfour}
\affiliation{SKA-SA, Cape Town, South Africa}

\author{Adam P. Beardsley}
\affiliation{School of Earth and Space Exploration, Arizona State University, Tempe, AZ}

\author{Tashalee S. Billings}
\affiliation{Department of Physics and Astronomy, University of Pennsylvania, Philadelphia, PA}

\author{Judd D. Bowman}
\affiliation{School of Earth and Space Exploration, Arizona State University, Tempe, AZ}

\author{Richard F. Bradley}
\affiliation{National Radio Astronomy Observatory, Charlottesville, VA}

\author{Philip Bull}
\affiliation{School of Physics \& Astronomy, Queen Mary University of London, London, UK}

\author{Jacob  Burba}
\affiliation{Department of Physics, Brown University, Providence, RI}

\author{Steven  Carey}
\affiliation{Cavendish Astrophysics, University of Cambridge, Cambridge, UK}

\author{Carina  Cheng}
\affiliation{Department of Astronomy, University of California, Berkeley, CA}

\author{David R. DeBoer}
\affiliation{Department of Astronomy, University of California, Berkeley, CA}

\author{Matt  Dexter}
\affiliation{Department of Astronomy, University of California, Berkeley, CA}

\author{Eloy  de~Lera~Acedo}
\affiliation{Cavendish Astrophysics, University of Cambridge, Cambridge, UK}

\author{John  Ely}
\affiliation{Cavendish Astrophysics, University of Cambridge, Cambridge, UK}

\author{Aaron  Ewall-Wice}
\affiliation{Department of Physics, Massachusetts Institute of Technology, Cambridge, MA}

\author{Nicolas  Fagnoni}
\affiliation{Cavendish Astrophysics, University of Cambridge, Cambridge, UK}

\author{Randall  Fritz}
\affiliation{SKA-SA, Cape Town, South Africa}

\author{Steve R. Furlanetto}
\affiliation{Department of Physics and Astronomy, University of California, Los Angeles, CA}

\author{Kingsley  Gale-Sides}
\affiliation{Cavendish Astrophysics, University of Cambridge, Cambridge, UK}

\author{Brian  Glendenning}
\affiliation{National Radio Astronomy Observatory, Socorro, NM}

\author{Deepthi  Gorthi}
\affiliation{Department of Astronomy, University of California, Berkeley, CA}

\author{Bradley  Greig}
\affiliation{School of Physics, University of Melbourne, Parkville, VIC 3010, Australia}

\author{Jasper  Grobbelaar}
\affiliation{SKA-SA, Cape Town, South Africa}

\author{Ziyaad  Halday}
\affiliation{SKA-SA, Cape Town, South Africa}

\author{Bryna J. Hazelton}
\affiliation{Department of Physics, University of Washington, Seattle, WA}
\affiliation{eScience Institute, University of Washington, Seattle, WA}

\author{Jacqueline N. Hewitt}
\affiliation{Department of Physics, Massachusetts Institute of Technology, Cambridge, MA}

\author{Jack  Hickish}
\affiliation{Department of Astronomy, University of California, Berkeley, CA}

\author{Daniel C. Jacobs}
\affiliation{School of Earth and Space Exploration, Arizona State University, Tempe, AZ}

\author{Austin  Julius}
\affiliation{SKA-SA, Cape Town, South Africa}

\author{Joshua  Kerrigan}
\affiliation{Department of Physics, Brown University, Providence, RI}

\author{Piyanat  Kittiwisit}
\affiliation{School of Earth and Space Exploration, Arizona State University, Tempe, AZ}

\author{Saul A. Kohn}
\affiliation{Department of Physics and Astronomy, University of Pennsylvania, Philadelphia, PA}

\author{Matthew  Kolopanis}
\affiliation{School of Earth and Space Exploration, Arizona State University, Tempe, AZ}

\author{Adam  Lanman}
\affiliation{Department of Physics, Brown University, Providence, RI}

\author{Paul  La~Plante}
\affiliation{Department of Physics and Astronomy, University of Pennsylvania, Philadelphia, PA}

\author{Telalo  Lekalake}
\affiliation{SKA-SA, Cape Town, South Africa}

\author{Adrian  Liu}
\affiliation{Department of Astronomy, University of California, Berkeley, CA}
\affiliation{Hubble Fellow}
\affiliation{Department of Physics and McGill Space Institute, McGill University, 3600 University Street, Montreal, QC H3A 2T8, Canada}

\author{David  MacMahon}
\affiliation{Department of Astronomy, University of California, Berkeley, CA}

\author{Lourence  Malan}
\affiliation{SKA-SA, Cape Town, South Africa}

\author{Cresshim  Malgas}
\affiliation{SKA-SA, Cape Town, South Africa}

\author{Matthys  Maree}
\affiliation{SKA-SA, Cape Town, South Africa}

\author{Zachary E. Martinot}
\affiliation{Department of Physics and Astronomy, University of Pennsylvania, Philadelphia, PA}

\author{Eunice  Matsetela}
\affiliation{SKA-SA, Cape Town, South Africa}

\author{Andrei  Mesinger}
\affiliation{Scuola Normale Superiore, 56126 Pisa, PI, Italy}

\author{Mathakane  Molewa}
\affiliation{SKA-SA, Cape Town, South Africa}

\author{Miguel F. Morales}
\affiliation{Department of Physics, University of Washington, Seattle, WA}

\author{Tshegofalang  Mosiane}
\affiliation{SKA-SA, Cape Town, South Africa}

\author{Steven G. Murray}
\affiliation{School of Earth and Space Exploration, Arizona State University, Tempe, AZ}

\author{Abraham R. Neben}
\affiliation{Department of Physics, Massachusetts Institute of Technology, Cambridge, MA}

\author{Bojan  Nikolic}
\affiliation{Cavendish Astrophysics, University of Cambridge, Cambridge, UK}

\author{Chuneeta D. Nunhokee}
\affiliation{Department of Astronomy, University of California, Berkeley, CA}

\author{Nipanjana  Patra}
\affiliation{Department of Astronomy, University of California, Berkeley, CA}

\author{Samantha  Pieterse}
\affiliation{SKA-SA, Cape Town, South Africa}

\author{Jonathan C. Pober}
\affiliation{Department of Physics, Brown University, Providence, RI}

\author{Nima  Razavi-Ghods}
\affiliation{Cavendish Astrophysics, University of Cambridge, Cambridge, UK}

\author{Jon  Ringuette}
\affiliation{Department of Physics, University of Washington, Seattle, WA}

\author{James  Robnett}
\affiliation{National Radio Astronomy Observatory, Socorro, NM}

\author{Kathryn  Rosie}
\affiliation{SKA-SA, Cape Town, South Africa}

\author{Peter  Sims}
\affiliation{Department of Physics, Brown University, Providence, RI}

\author{Craig  Smith}
\affiliation{SKA-SA, Cape Town, South Africa}

\author{Angelo  Syce}
\affiliation{SKA-SA, Cape Town, South Africa}

\author{Nithyanandan  Thyagarajan}
\affiliation{School of Earth and Space Exploration, Arizona State University, Tempe, AZ}
\affiliation{National Radio Astronomy Observatory, Socorro, NM}

\author{Peter K.~G. Williams}
\affiliation{Center for Astrophysics, Harvard \& Smithsonian, 60 Garden St., Cambridge, MA}
\affiliation{American Astronomical Society, 1667 K Street NW, Suite 800, Washington, DC 20006}

\author{Haoxuan  Zheng}
\affiliation{Department of Physics, Massachusetts Institute of Technology, Cambridge, MA}

\begin{abstract}
We discuss absolute calibration strategies for Phase I of the Hydrogen Epoch of Reionization Array (HERA), which aims to measure the cosmological \tocm signal from the Epoch of Reionization (EoR).
HERA is a drift-scan array with a $10^\circ$ wide field of view, meaning bright, well-characterized point source transits are scarce.
This, combined with HERA's redundant sampling of the $uv$ plane and the modest angular resolution of the Phase I instrument, make traditional sky-based and self-calibration techniques difficult to implement with high dynamic range.
Nonetheless, in this work we demonstrate calibration for HERA using point source catalogues and electromagnetic simulations of its primary beam.
We show that unmodeled diffuse flux and instrumental contaminants can corrupt the gain solutions, and present a gain smoothing approach for mitigating their impact on the 21\,cm power spectrum.
We also demonstrate a hybrid sky and redundant calibration scheme and compare it to pure sky-based calibration, showing only a marginal improvement to the gain solutions at intermediate delay scales.
Our work suggests that the HERA Phase I system can be well-calibrated for a foreground-avoidance power spectrum estimator by applying direction-independent gains with a small set of degrees of freedom across the frequency and time axes.
\end{abstract}

%%%%%%%%%%%%%%%%%%%%%%%%%%%%%%
%%%%%%%%%%%% Introduction %%%%%%%%%%%%
%%%%%%%%%%%%%%%%%%%%%%%%%%%%%%

\section{Introduction}
\label{sec:intro}	
The Hydrogen Epoch of Reionization Array\footnote{\url{http://reionization.org/}} \citep[HERA;][]{DeBoer2017} is a targeted, radio interferometric experiment that aims to measure the cosmological 21\,cm spin-flip emission from primordial hydrogen in the intergalactic medium (IGM) at Cosmic Dawn.
One of the last frontiers of cosmology and high redshift astrophysics, the Cosmic Dawn marks the era when the first stars, black holes, and galaxies formed and interacted with the surrounding IGM.
Eventually, these sources heated and re-ionized the majority of the neutral hydrogen in the IGM, in an event known as the Epoch of Reionization (EoR).
A number of questions remain about when and how the Cosmic Dawn and EoR occurred, which are crucial to our broader understanding of galaxy and large-scale structure formation.
For reviews, see \citet{Morales2010, Mesinger2016b, Liu2019}.

One of the only direct probes of the IGM throughout the entirety of Cosmic Dawn is neutral hydrogen's 21\,cm transition, which at redshifts of $z\sim10$ appears in the low-frequency radio band around 150 MHz.
Over the past decade, first-generation 21\,cm EoR experiments like the Donald C. Backer Precision Array for Probing the Epoch of Reionization \citep[PAPER;][]{Parsons2014, Jacobs2015, Cheng2018, Kolopanis2019}, the Murchison Widefield Array \citep[MWA;][]{Tingay2013, Dillon2014, Ewall-Wice2016b, Beardsley2016, Barry2019b, Li2019}, the Low Frequency Array \citep[LOFAR;][]{vanHaarlem2013, Patil2017, Gehlot2018b}, the Giant Metre Wave Radio Telescope \citep[GMRT;][]{Paciga2013}, 
and the Long Wavelength Array \citep[LWA;][]{Eastwood2019}
 have set increasing stringent limits on the Cosmic Dawn 21\,cm power spectrum.
Meanwhile, global signal experiments have placed constraints on the 21\,cm monopole \citep{Bernardi2016, Singh2017}, with a reported first detection of the signal at Cosmic Dawn from the Experiment to Detect the Global EoR Signature \citep[EDGES;][]{Bowman2018}.
21\,cm experiments face the challenge of separating-out the weak cosmological signal from galactic and extra-galactic foreground emission that is many orders of magnitude brighter.
However, the 21\,cm signal is expected to be highly spectrally variant due to inhomogeneities in the density, ionization state and temperature of the IGM along the line-of-sight, while non-thermal foreground emission is expected to be spectrally smooth.
This provides a means for separating foreground emission from the desired cosmological signal.
However, even small instrumental effects can distort these foregrounds and contaminate the region in Fourier space occupied nominally only by the EoR signal and thermal noise, known as the EoR window \citep{Morales2012}.
High dynamic range instrumental gain calibration is therefore critical to 21\,cm science.

Per-antenna gain calibration is the task of solving for a single complex number per antenna and feed polarization (as a function of both time and frequency) that best satisfies the antenna-based calibration equation for a visibility $V_{ij}$ defined between antenna $i$ and antenna $j$,
\begin{align}
\label{eqn:ME}
V_{ij}^{\rm measured}(\nu, t) = V_{ij}^{\rm true}(\nu, t)g_i(\nu, t)g_j^\ast(\nu, t),
\end{align}
where $V_{ij}^{\rm measured}$ is the raw data, $V_{ij}^{\rm true}$ is the true visibility that would be measured by an uncorrupted instrument, and $g_i$ and $g_j$ are the instrumental gains for antenna $i$ and $j$, respectively \citep{Hamaker1996}.
Recent work has shown how incomplete models in sky-based calibration \citep{Barry2016, Ewall-Wice2017, Byrne2019} and non-redundancies in redundant calibration \citep{Joseph2018, Orosz2019} can lead to gain calibration errors that contaminate the EoR window.
Foreground and instrument simulations for HERA indicate that the fiducial EoR signal at $k\sim0.2\ h\ {\rm Mpc}^{-1}$ is expected to be roughly $10^{5}$ times weaker than the peak foreground amplitude at $k\sim0\ h\ {\rm Mpc}^{-1}$ in the visibility \citep{Thyagarajan2016}.
Because gain calibration is multiplicative in frequency it can equivalently be thought of as a convolution in delay space, the Fourier dual of frequency.
This means that each antenna's gain kernel, or the gain's footprint in delay space, must be nominally suppressed by at least five orders of magnitude at delay scales of $\tau\gtrsim400$ ns (400 ns equals $k_\parallel=0.2\ h\ {\rm Mpc}^{-1}$ at $z=10$ or $\nu\sim130$ MHz for the 21\,cm line).
In this case we have chosen to represent the gains as \emph{direction-independent}, which is the component of gains we are concerned with in this work, although much work has been devoted to \emph{direction-dependent} gain calibration \citep[e.g.][]{Bhatnagar2008, Intema2014}.

HERA was deployed in two stages, Phase I and Phase II.
Phase I observed from 2017 - 2018 while only a section of the array was built and used front-end signal chains from the PAPER experiment.
Phase II is currently under construction towards a build-out of 350 antennas and will be equipped with completely new front-end hardware (HERA Collaboration in prep).
The work in this paper uses only Phase I observations (\autoref{sec:observations}).
HERA is a drift-scan array, meaning it is built into the ground and cannot physically point its antennas on the sky.
With its 10$^\circ$ degree field of view (FoV), the number of bright and well-characterized point sources that transit on any given night are limited.
Furthermore, the highly redundant $uv$ sampling and relatively short baselines of the HERA Phase I configuration make implementing self-calibration to high dynamic ranges difficult.
Nonetheless, we outline a strategy for sky-based calibration of HERA Phase I using point sources from the MWA's GLEAM catalogue \citep{Hurley-Walker2017} and electromagnetic simulations of HERA's primary beam \citep{Fagnoni2019}.
We show this does a fairly good job at bringing the data in-line with the adopted model, and use it to characterize the frequency and time stability of the gains.
Importantly, we also show that performing antenna-based calibration in the presence of non-antenna-based systematics can contaminate systematic-free visibilities.
We discuss the impact this has on the data and the \tocm power spectrum, and demonstrate gain smoothing procedures to mitigate this and other gain errors introduced in the process of calibration.

Redundant calibration has been hailed as a powerful alternative calibration strategy for 21\,cm experiments that sidesteps some of the requirements of sky-based calibration \citep{Liu2010, Zheng2014}.
However, redundant calibration still needs a sky model to pin down certain degenerate parameters it cannot solve for \citep{Dillon2018, Li2018, Joseph2018, Byrne2019}.
In this work, we explore hybrid redundant-absolute calibration strategies using the \texttt{hera\_cal} package.\footnote{\url{https://github.com/HERA-Team/hera_cal}}
Applying them to HERA Phase I, we show that redundant calibration seems to mitigate some errors associated with sky-based calibration, however, it also has its own set of uncertainties due to inherent non-redundancies that need to be mitigated.
For low delay modes in the gains, we find that redundant and sky calibration yield very similar results.

In this work, we use the term \emph{absolute calibration} to refer to the components of the full antenna-based gains that are constant across the array (note these are still frequency dependent).
One example of this is the average antenna gain amplitude, which sets the overall flux scale of the data.
Indeed, these are exactly the terms that are degenerate in redundant calibration.
In sky-based calibration these terms are automatically solved for, which can therefore be thought of as a form of absolute calibration.

The structure of this paper is as follows:
in \S2 we detail the observations used in this analysis.
In \S3 we describe our methodology for sky-based calibration of HERA.
In \S4 we characterize the time and frequency stability of the gain solutions.
In \S5 we synthesize redundant and absolute calibration and compare them to traditional sky-based calibration.
In \S6 we calibrate the data and investigate foreground contamination in the power spectrum, and in \S7 we summarize our results.

%%%%%%%%%%%%%%%%%%%%%%%%%%%%%%%
%%%%%%%%%%%% Observations %%%%%%%%%%%%
%%%%%%%%%%%%%%%%%%%%%%%%%%%%%%%

\section{Observations}
\label{sec:observations}
The data used in this work were taken with the HERA Phase I instrument \citep{DeBoer2017} in a 56-element configuration on Dec. 10, 2017 (Julian Date 2458098).
HERA is located in the Karoo Desert, South Africa, at the South African Karoo Radio Astronomy Reserve.
Data were taken in drift-scan mode for roughly 12 hours per night starting at 5pm South African Standard Time, of which roughly 9 hours are deemed good quality data when the Sun is below the horizon.

\begin{figure}
\label{fig:hera_ant}
\centering
\includegraphics[scale=0.12]{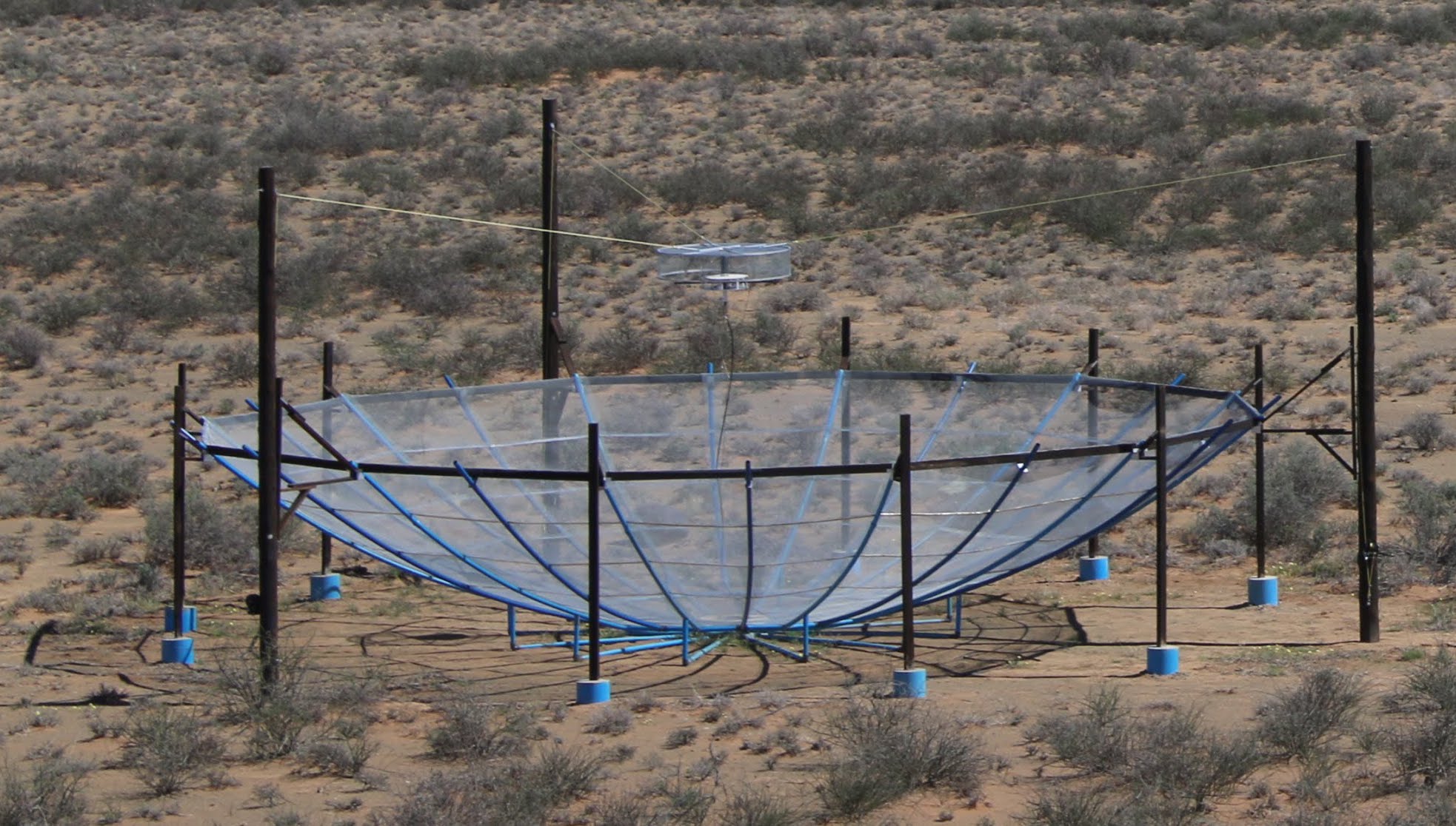}
\caption{A single HERA antenna in the field with a cross-dipole feed surrounded by a cage hoisted to the antenna's focal point. Image courtesy of Kathryn Rosie.}
\end{figure}

The Phase I instrument repurposed many of the older PAPER experiment components, including its signal chains, correlator, feeds and front-end modules (FEM), and attached them to newly designed HERA antennas.
The HERA antenna (\autoref{fig:hera_ant}) is a 14-meter dish with an optimized version of the dual linear polarization PAPER feed and FEM hoisted 4.9 meters to its focal height.
The optimized feed and dish were designed to minimize reflections within the antenna, and thus limit excess chromaticity induced by the signal chain \citep{Neben2016, Thyagarajan2016, Ewall-Wice2016c, Patra2018}.
From the FEM, which houses an initial stage of amplification, the analog chain consists of a 150-meter coaxial cable connected to a node unit in the field where the signals are fed through a post amplification stage (PAM) and a filtering stage.
From there, the signals travel through another 20-meter coaxial cable to a container where they are digitized, Fourier transformed and then cross-multiplied with all other antenna and linear polarization streams.
Additional observational parameters are detailed in \autoref{tab:hera_obs}.

\begin{figure*}
\label{fig:uvplot}
\centering
\includegraphics[scale=0.5]{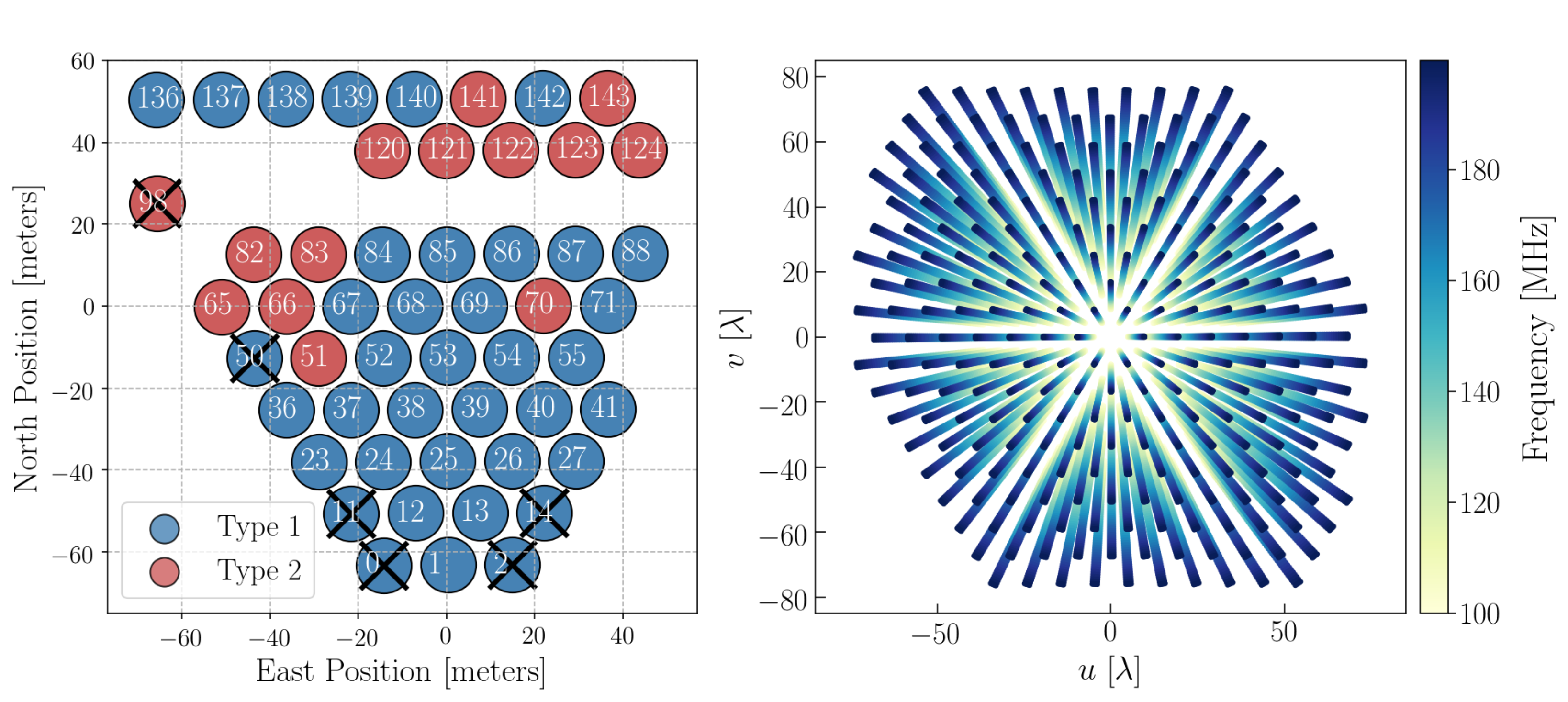}
\caption{{\bfseries Left}: The HERA Phase I array layout with 56 connected antennas and 50 operational antennas.
Antennas determined to be problematic are marked with crosses.
{\bfseries Right}: The corresponding $uv$ sampling of the array over a 10-minute time window and a frequency range of 100 -- 200 MHz, highlighting HERA's highly redundant $uv$ sampling. The color gradient represents independent $uv$ samples throughout the total bandwidth.}
\end{figure*}

Not all of the PAPER signal chains could be salvaged for the HERA Phase I instrument.
As a temporary stopgap, additional FEMs, cables and PAMs were manufactured for Phase I data collection.
We refer to the new set of signal chains as ``Type 1'' and the old set of signal chains as ``Type 2,''
which are colored blue and red in \autoref{fig:uvplot}, respectively.
The transmission properties of the signal chains are studied in more detail in \citet{Kern2019b}.
For more details on the HERA Phase I signal chain and electronics, we refer the reader to \citet{Parsons2010, DeBoer2017, Fagnoni2019}.

Before calibration, the data are pre-processed with part of the HERA analysis pipeline.
Specifically, faulty antennas are identified and flagged at a quality metrics stage (crosses in \autoref{fig:uvplot}) and radio frequency interference (RFI) is excised from the data using median filtering and a watershed algorithm \citep{Kerrigan2019}.
The data are written to disk in the Miriad file format post-correlation, which are then converted to \texttt{UVFITS} using the \texttt{pyuvdata} software \citep{Hazelton2017} and imported to CASA Measurement Sets via CASA's \texttt{importuvfits} task.

\begin{deluxetable}{lc} 
\tabletypesize{\footnotesize} 
\tablewidth{0pt} 
\tablecaption{
HERA Observation Parameters
\label{tab:hera_obs}
}
\tablehead{Parameter & Value}
\startdata 
Array Configuration & Phase I \\[.1cm]
Number of Antennas & 56\\[.1cm]
Array Coordinates  & -30.7$^\circ$ S, 21.4$^\circ$E  \\[.1cm]
Observing Mode & drift-scan \\[.1cm]
Correlator Integration & 10.7 seconds \\[.1cm]
Frequency Range & 100 - 200 MHz \\[.1cm]
Channel Width & 97.65 kHz \\[.1cm]
Dish Diameter & 14 meter \\[.1cm]
Feed Type & dual polarization X \& Y dipoles \\[.1cm]
Visibility Polarizations & XX, XY, YX, YY \\[.1cm]
Shortest / Longest Baseline & 14.6 / 139.3 meters \\[.1cm]
Observation Dates & December 10, 2017 \\[.1cm]
\enddata 
%\vspace{-.3cm} 
\tablecomments{For the 2017--2018 observation, the HERA correlator used the convention that the X dipole points East-West while the Y dipole points North-South, which is not the standard \citet{Hamaker1996b} definition that assumes the opposite.}
\end{deluxetable}

%%%%%%%%%%%%%%%%%%%%%%%%%%%%%%%
%%%%%%%%%%%% Calibration %%%%%%%%%%%%%
%%%%%%%%%%%%%%%%%%%%%%%%%%%%%%%
\section{Sky-Based Calibration}
\label{sec:skycal}

Standard sky-based calibration is typically done by choosing a bright, well-characterized point source for the model visibilities.
This is made difficult for HERA because it is a drift-scan array, meaning it cannot be pointed to an arbitrary location on the sky.
Furthermore, the larger collecting area provided by a dish, as opposed to a lone dipole, means HERA's primary beam response is more compact on the sky compared to other experiments like PAPER or the MWA:
at 150 MHz, HERA's primary beam FWHM is roughly $10^\circ$, compared to roughly $45^\circ$ for the PAPER experiment.
This means that the number of bright, well-characterized radio sources that transit our field of view is low.
In fact, not a single point source within 5$^\circ$ of HERA's declination exceeds 20 Jy in flux density in the cold part of the radio sky (far from the galactic plane).
Implementing self-calibration to high dynamic range is also difficult for HERA given its highly redundant sampling of the $uv$ plane, making HERA's narrow-band grating lobes very severe.
This is compounded by the poor angular resolution of the Phase I instrument, making it quickly confusion noise limited (\autoref{fig:uvplot}).
Redundant calibration somewhat skirts the problem of an inadequate sky model, and indeed exploiting the power of redundant calibration was a motivating factor behind HERA's redundant design \citep{Dillon2016}.
However, redundant calibration operates only within a specific subspace of the full antenna-based calibration equations, meaning a model of the sky is still fundamentally needed to fill in the few remaining degenerate modes \citep{Liu2010, Zheng2014, Dillon2018, Li2018, Byrne2019, Dillon2019}.
We discuss this in more detail for HERA in \autoref{sec:redcal}.

For power spectrum estimators that do not attempt to subtract the dominant foreground emission in the data (at the expense of losing low $k$ modes), the stringent requirement of high dynamic range source modeling is relaxed because we are not interested in recovering modes inherently occupied by foreground emission.
Hybrid techniques also exist, which try to reap the benefits of both foreground removal and avoidance \citep{Kerrigan2018}.
For foreground avoidance estimators, a path towards achieving deep, noise-limited power spectrum integrations at intermediate spatial modes of $k\gtrsim0.2\ h\ {\rm Mpc}^{-1}$ with a calibration derived from the sky may be possible even with the challenges faced by the HERA Phase I instrument.
In this section we describe a sky-based calibration strategy for HERA using custom pipelines for calibration and imaging \footnote{\url{https://github.com/HERA-Team/casa_imaging}} built around the Common Astronomy Software Applications \citep[CASA;][]{McMullin2007} package.
We start by discussing the construction of our flux density model, and then describe our calibration methodology and its validation via imaging and source extraction.

\subsection{Building a Sky Model}
\begin{figure*}
\label{fig:gsm_fields}
\centering
\includegraphics[scale=0.63]{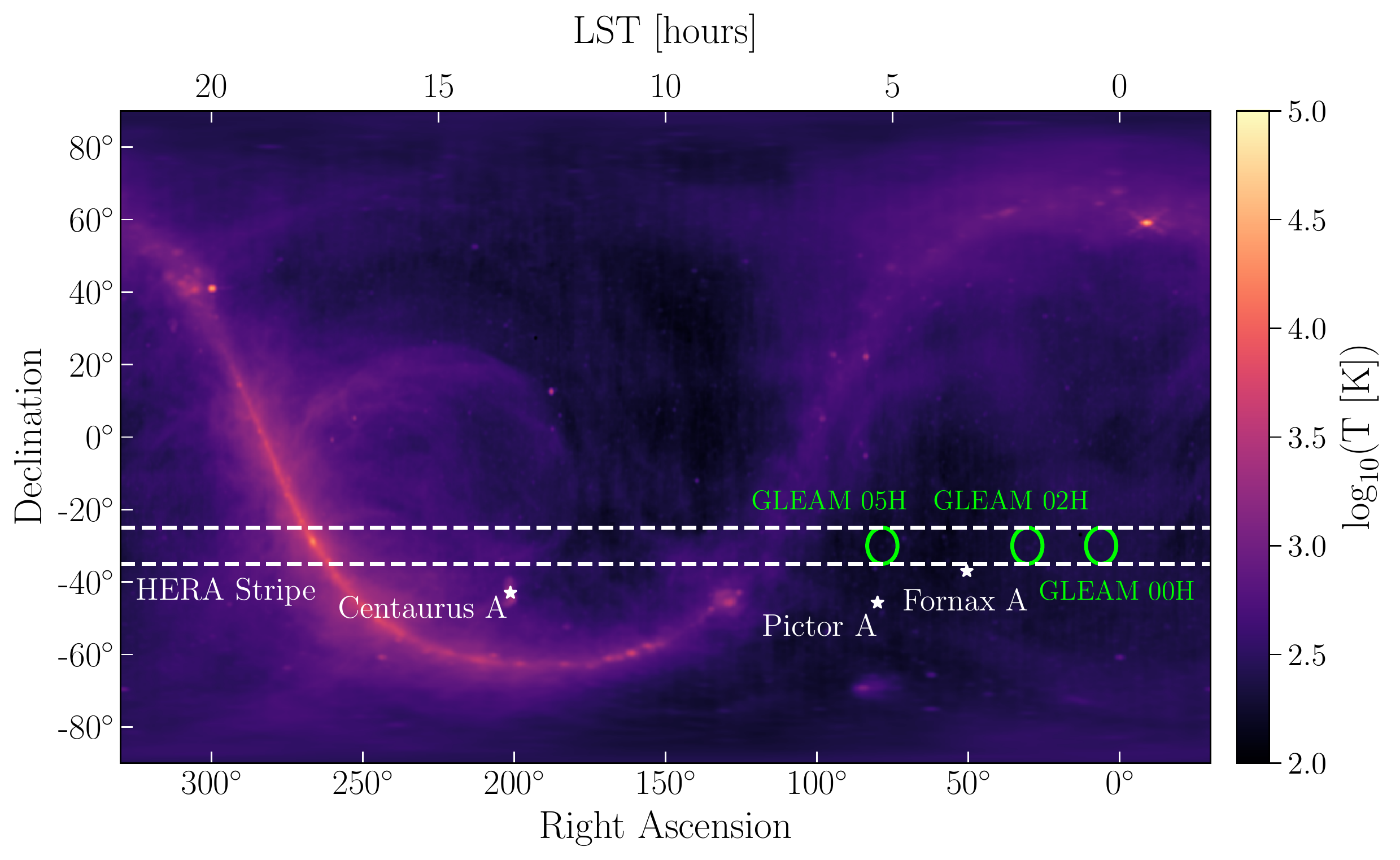}
\caption{The radio sky at 150 MHz from the GSM \citep{Oliveira2008}, showing the bright galactic and extra-galactic foregrounds that stand in the way of cosmological \tocm experiments.
The HERA stripe is shown in dashed lines centered at HERA's declination of -30.7$^\circ$ with a width of $10^\circ$, which is the FWHM of the primary beam at 150 MHz.
The three fields identified as ideal calibration fields are shown in green circles, and some bright extended sources in the vicinity are marked as stars.
 }
\end{figure*}

Our ideal model for sky-based calibration would involve a single, bright point source located at the pointing center of the field-of-view (FoV).
Because HERA is a drift-scan array, this means our ideal calibrator would be located at $\delta\sim-30.7^\circ$ and would transit zenith at some point in the night.
Ideally this calibrator would be so bright that other off-axis point sources or diffuse emission would contribute a vastly subdominant component of the measured visibilities.
Unfortunately this is not the case for HERA, so we are forced to make compromises.
\autoref{fig:gsm_fields} is a map of radio foregrounds at 150 MHz from the Global Sky Model \citep{Oliveira2008} and shows the HERA stripe (white-dashed), which denotes the track of the FWHM of HERA's primary beam ($10^\circ$ at 150 MHz).
We see that the HERA stripe covers a fairly small part of the sky, demonstrating how limited we are in the amount of sky available for identifying bright calibrators.

To select the best calibration field given our limitations, we can identify some key criteria that a good field should satisfy.
The first criterion is that the field should have most of its radio emission contained in the main-lobe of the primary beam.
Off-axis sources located in the far side-lobes of the primary beam are troublesome because primary beam side-lobes are hard to model accurately.
One workaround is to peel these sources from the visibilities before calibration \citep[e.g.][]{Hurley-Walker2017, Eastwood2019} but that requires one to image them at a fine frequency resolution to capture primary beam chromaticity and also with high dynamic range, which as stated is challenging for HERA Phase I.
Additionally, we want our direction-independent calibration to be representative of the instrument response \emph{at zenith}, because that is where most of the measured EoR signal comes from.
Said another way, we do not want our direction-independent calibration to soak up structure from direction-dependent effects introduced by off-axis sources.
One example of this is diffuse emission coming from the plane of the galaxy, which extends across the entire FoV when it transits.

The second criterion for a good calibration field is that it should have sources that are well characterized at the observing frequencies.
Furthermore, it should have a relatively bright source very close to the FoV pointing center so that we can confirm via imaging that our calibration at zenith yields a good match to the input model.
Such a source can also be useful for empirically characterizing the primary beam response with drift-scan source tracks \citep{Pober2012, Eastwood2018, Nunhokee2019}.

Recently, the MWA constructed the GLEAM point source catalogue \citep{Hurley-Walker2017} from a deep, low-frequency survey spanning the Southern Hemisphere, overlapping with the HERA stripe.
We searched the GLEAM catalogue for all point sources within $2.5^\circ$ degrees of $\delta=-30.7^\circ$ with a flux density above 15 Jy at 150 MHz, located in the cold part of the radio sky (LST $<$ 6 hours).
We find three such sources in the GLEAM catalog, J0024-2929 at 0 hours LST, J0200-3053 at 2 hours LST and J0455-3006 at 5 hours LST.
Their positions, flux densities, and spectral indices are reported in \autoref{tab:calibrators}.
\citet{Jacobs2016} performed a similar exercise with the TGSS ADR catalog \citep{Intema2017}.
They also find J0200-3053 as a possible calibrator, but do not identify the other two sources we quote from the GLEAM catalog.
For the shared source, the quoted values between the GLEAM and TGSS ADR catalogs agree to within 15\%, which is roughly in-line with the overall accuracy of the survey flux scales.
The green circles in \autoref{fig:gsm_fields} are centered on each of these three calibration fields, and have diameter equal to the $10^\circ$ FWHM of the HERA primary beam at 150 MHz.
Stars mark the location of the nearby bright, extended sources like Pictor A and Fornax A.

\begin{deluxetable}{lccccc}[h!]
\tabletypesize{\footnotesize} 
\tablewidth{0pt} 
\tablecaption{
HERA Calibrator Candidates from GLEAM
\label{tab:calibrators}
}
\tablehead{Name & RA (J2000) & Dec (J2000) & $S_{\rm peak}$ & $S_{\rm int}$ & $\alpha$}
\startdata 
J0024-2929 & 6.126 & -29.48 & 16.45 & 16.10 & -0.867 \\[.1cm]
J0200-3053 & 30.05 & -30.89 & 19.50 & 17.95 & -0.863 \\[.1cm]
J0455-3006 & 73.81 & -30.11 & 16.34 & 17.11 & -0.781 \\[.1cm]
\enddata 
%\vspace{-.3cm} 
\tablecomments{All GLEAM \citep{Hurley-Walker2017} sources above 15 Jy, with LST $<$ 6 hours and $-33.2 < \delta < -28.2$. Equatorial coordinates are in degrees, flux densities are in Jy at 151 MHz and $\alpha$ is the spectral index anchored at 151 MHz.}
\end{deluxetable}

\begin{figure*}
\label{fig:flux_model}
\centering
\includegraphics[scale=0.60]{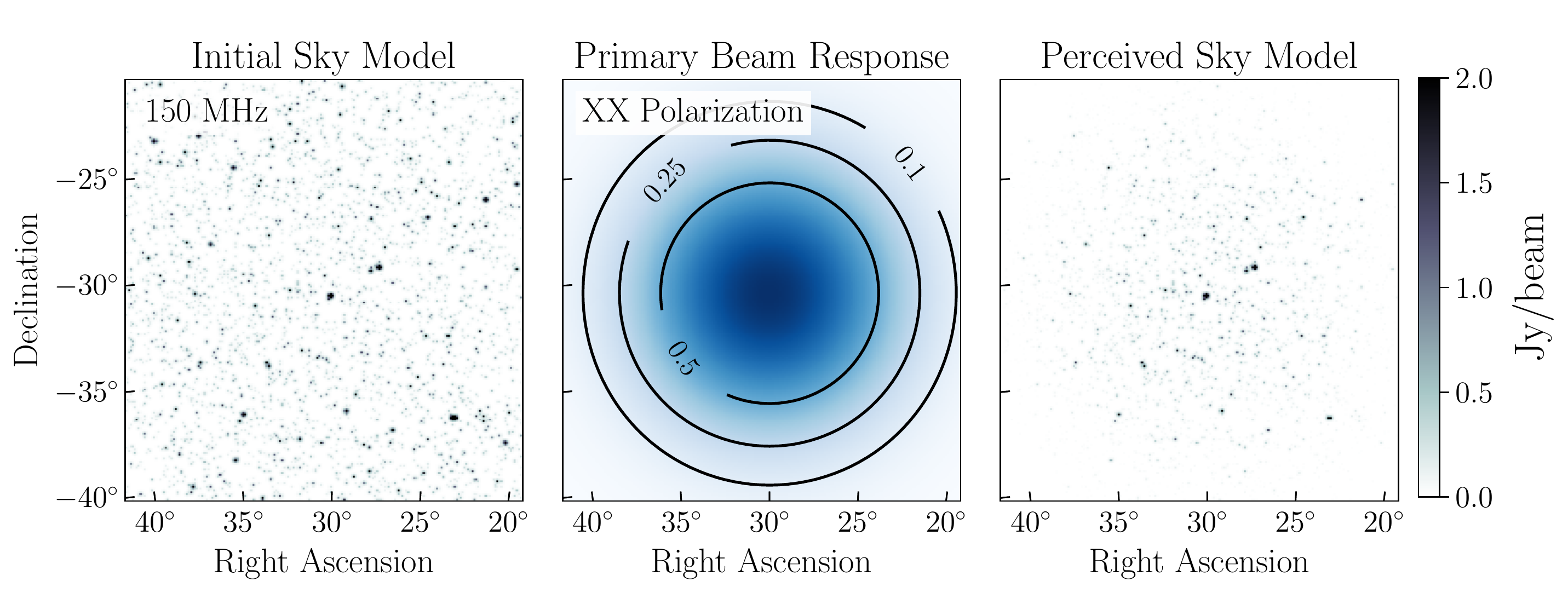}
\caption{Construction of the GLEAM-02H field sky model for calibration at 150 MHz.
Each frequency channel in the model is constructed independently in the same manner.
{\bfseries Left}: All GLEAM point sources in Stokes I polarization above 0.1 Jy within 20$^\circ$ of the pointing center.
In this figure, the point sources have been convolved with a narrow 2D Gaussian merely for visual clarity.
{\bfseries Center}: The peak-normalized primary beam response for the XX instrumental linear polarization at 150 MHz \citep{Fagnoni2019}.
{\bfseries Right}: The Stokes I model multiplied by the XX primary beam response yields a perceived flux density model that is then converted into visibilities for calibration.}
\end{figure*}

Even though a $\sim$20 Jy primary calibrator source exists at the pointing center of each field, they themselves make up only a fraction of the total flux density measured by the instrument at those LSTs.
For short baselines the dominant sky component is diffuse galactic emission, while longer baselines are dominated by point sources spread across the FoV.
Although models of the diffuse galactic emission exist \citep{Oliveira2008, Zheng2016} they are only accurate at the $\sim$15\% and furthermore extend across the entire FoV, filling the hard-to-model sidelobes.
At the moment, we only use point sources in our flux density model and cut short baselines ($<40$ meters) that have significant amounts of diffuse foreground emission.
Our starting model for each field is made up of all GLEAM point sources down to 0.1 Jy in flux density extending 20$^\circ$ in radius from the pointing center, which typically results in $\sim$10,000 sources in the flux density model.
We take the GLEAM-reported flux density of each point source at 151 MHz and their spectral index and insert them into a CASA component list.
All sources are assumed to be unpolarized and their fluxes are inserted purely as Stokes I.
For GLEAM sources without a spectral index, we take the reported flux density of the source at 122, 130, 143, 151, 158, 166, and 174 MHz and fit our own spectral index.
After constructing a component list with all of the relevant GLEAM sources we make a 1024-channel spectral cube image of the component list with the CASA \texttt{Image.modify} task, matching the channelization of HERA data, and export it to \texttt{FITS} format.
The image has a pixel resolution of 300 arcseconds, which is 6 times smaller than the synthesized beam FWHM of $\sim$0.5 degrees.

Note that the GLEAM catalogue does not include bright, extended sources like Fornax A and Pictor A.
As shown in \autoref{fig:gsm_fields}, the calibration fields are chosen such that these sources are heavily attenuated by the primary beam, but even still these sources can be seen at the level of a few Jansky for the 02-hour and 05-hour fields, for example.
Fornax A and Pictor A can be included in the component list model for the GLEAM-02H and GLEAM-05H fields, respectively, by adopting point source models with spectral indices informed by recent low-frequency studies \citep{Jacobs2013, McKinley2015}.
Although these sources have a non-zero angular extent to them, for HERA Phase I angular resolutions a point source model is a fair approximation.

Next we incorporate the effects of the direction and frequency-dependent antenna primary beam response to create a perceived flux density model.
We use an electromagnetic simulation of the HERA primary beam from \citet{Fagnoni2019}, which includes effects from the dish and feed.
That work also explored the effects of mutual coupling on the primary beam response given an element embedded in the array, finding second-order effects on the beam response near the horizon at the level of $10^{-2}$ in power.
Empirical studies by \citet{Kern2019b} find similar levels of mutual coupling in the data, and present post-calibration methods for mitigating their effects.
In this work we only use the \citet{Fagnoni2019} beam model of the antenna and feed, and defer using the embedded element pattern in calibration for future work.
Each linear dipole in the feed, X and Y, is assigned its own beam model, where one is simply a $90^\circ$ rotation of the other.
The beams are peak-normalized at boresight independently at each frequency, and we then multiply the beam response at each pixel on the sky separately for the X and Y dipoles.
This results in two spectral cubes, one for both the XX and YY instrumental visibility polarizations, which constitutes our perceived model.
In this work we do not construct models for the cross-polarized XY and YX visibilities as we will not perform polarization calibration, although this can be done with polarized beam models \citep{Martinot2018}.
\autoref{fig:flux_model} demonstrates this for the GLEAM 02H field in XX instrumental polarization, showing the initial sources (left), the XX primary beam response (or the squared X-dipole response) at 150 MHz (center), and the product of the two (right).
Lastly, the model cubes are transformed from the image to the $uv$ domain via CASA's \texttt{ft} task and are inserted into the model column of the Measurement Sets for calibration.

\subsection{Calibration}
Next we will describe our approach for deriving complex, direction-independent antenna gains with CASA.
For simplicity, we will focus our discussion specifically to the GLEAM-02H field, but calibration on any other field would follow the same procedures outlined below.
As noted, the data are first processed for faulty antennas and RFI flagging by the HERA analysis pipeline.
We then take five minutes of drift-scan data centered at the transit of the primary calibrator, apply a fringe-stop phasing to the transit LST and then time-average the data.
Averaging five minutes of data allows us to increase the signal-to-noise ratio (SNR) of the derived gains and is still a fairly short time interval compared to the FWHM primary beam crossing time (at 150 MHz) of $\sim$46 minutes, at which point sky source decorrelation will begin to be a problem.
Due to the inherent stability of the drift-scan observing mode, we do not expect the gains to vary substantially over such short time scales (although see \autoref{sec:temporal_response} for higher-order effects).

Before proceeding with calibration, we enact a minimum baseline cut such that all baselines shorter than 40 meters ($\sim20\lambda$) are excluded,
leaving 65\% of the visibilities for calibration.
HERA's shortest baselines are most sensitive to the diffuse galactic emission that is not included in our point source model.
After experimenting with various baseline cuts we find a 40-meter cut to be a good compromise between keeping as much data as possible for maximal gain SNR and eliminating diffuse foreground flux in the data that is not included in our model.

Our process for deriving antenna gains uses a series of standard routines in CASA.
Before each calibration step, we apply all previous calibration steps to the data on-the-fly.
The final calibration is then simply the product of all steps in our calibration chain.
We start by performing delay calibration using the \texttt{gaincal} task, which is done to calibrate out the cable delay of each antenna.
Next we perform mean-phase and mean-amplitude calibration (which consist of two numbers for each antenna-polarization across the entire bandwidth) also using the \texttt{gaincal} task.
This removes any residual phase offset after delay calibration and sets the overall flux scale of the data.
Up to now all calibration steps are smooth across frequency and therefore do not contain significant spectral structure.
Finally, we derive complex antenna bandpasses using the \texttt{bandpass} task, which solves each frequency channel independently from all others.
This last step has the possibility of introducing an arbitrary amount of spectral structure into the gain solutions and therefore deserves closer attention, which we revisit in \autoref{sec:gains}.

In this work we do not make any attempt to correct for effects due to the ionosphere.
This is less of a concern given the higher frequency range of 100 -- 200 MHz, as well as the limited angular resolution of the array and the fact that observations are only taken at night when the sun is below the horizon leading to calmer ionospheric conditions.
We also do not attempt to calibrate the relative phase between dipole polarizations in this work, which is difficult due to the dearth of bright polarized sky sources \citep{Moore2017, Lenc2017}, although this can still be partially constrained if we assert that the Stokes V visibilities be consistent with thermal noise \citep{Kohn2016}.
This is less of a concern because in this work we are mostly interested in the parallel-hand (i.e. XX and YY) dipole and Stokes I data products, which are not as sensitive to this term as the Stokes U \& V data products.
While previous work has shown that ionospheric leakage of point source foregrounds can in principle be significant \citep{Nunhokee2017}, ionospheric-induced leakage terms have also been shown to average down night-to-night \citep{Martinot2018}.
As we will show in \autoref{sec:imaging}, the amount of intrinsic polarization leakage observed in the data is quite small, even without performing any kind of polarization calibration.
Future HERA observations that i) extend below 100 MHz or ii) are interested in polarized data products will need to revisit these topics.
For an investigation into direction-dependent effects and polarization leakage from the HERA-19 commissioning array see \citet{Kohn2019}.

 \subsection{Imaging}
 \label{sec:imaging}
 
 \begin{figure*}
\label{fig:gleam02_MFS_XX}
\centering
\includegraphics[scale=0.65]{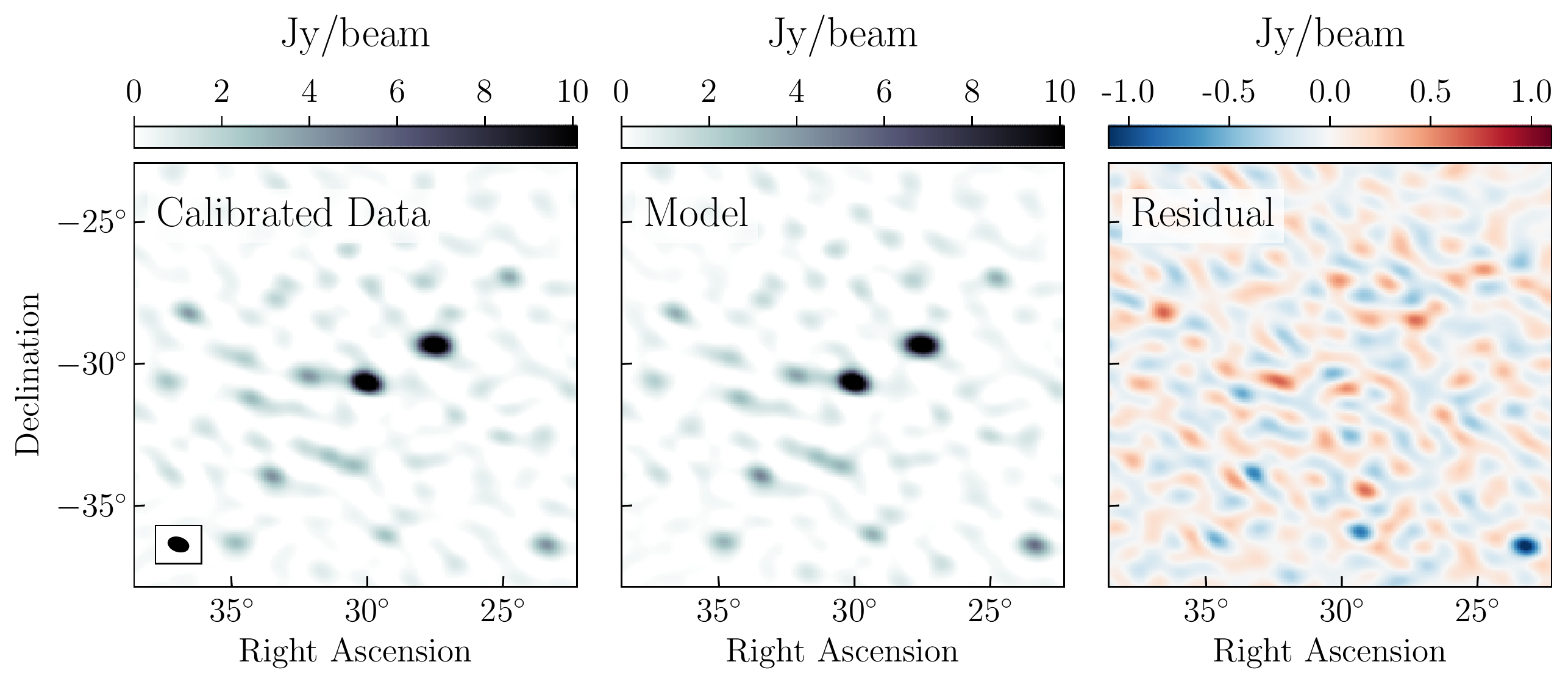}
\caption{Multi-frequency synthesis image of the GLEAM-02H field in XX polarization spanning 120 --- 180 MHz of the calibrated visibilities (left), model visibilities (center) and the residual visibilities (right).
Each image is CLEANed with the same parameters down to 0.5 Jy, with the restoring beam shown in the lower left.
The model and calibrated data show good agreement in the main lobe of the primary beam.
At larger zenith angles the residual image shows evidence for mis-calibration, likely due to primary beam errors.}
\end{figure*}

To test the fidelity of the calibration, we make multi-frequency synthesis (MFS) images of the calibrated data, the calibration model and their residual visibility as a visual assessment of their agreement.
The MFS images use five minutes of data and a 60 MHz bandwidth spanning 120 -- 180 MHz.
All images are made from only the baselines involved in the calibration ($|b| >$ 40 m), employ robust weighting with \texttt{robust = -1} and adopt the Hogbom deconvolution algorithm \citep{Hogbom1974} using the \texttt{tclean} task.
All images are CLEANed independently down to a threshold of 0.5 Jy in the polarization they are imaged in.
CLEAN masks are used around the brightest sources initially and then the CLEAN mask is opened up to the entire field.
We produce images in instrumental XX and YY polarization and also pseudo-Stokes I, Q, U \& V polarization.

The HERA array is not perfectly co-planar, which will introduce artifacts into wide-field images made with CASA.
This can be mitigated with W-projection \citep{Cornwell2008}, however, given the field of view and modest angular resolution of the Phase I array, we do not expect non-co-planar effects to generate an appreciable amount of error.
Therefore we do not perform W-projection in the process of imaging, which also reduces its overall computational cost.

\autoref{fig:gleam02_MFS_XX} shows the GLEAM-02H field in XX polarization and images of its calibrated data (left), model (center) and their residual visibility (right).
The size of the synthesized beam is shown in the lower left.
We see good agreement between the data and model down to a few percent.
The residual image appears noise-like in the main lobe, but further away from the pointing center we can begin to correlate point sources in the data with point sources in the residual.
This is a result of an improper perceived flux density model (either with the inherent source fluxes or, more likely, the adopted primary beam response).
This will introduce spectrally-dependent errors into the gain solutions at some level \citep{Barry2016, Ewall-Wice2017} which we explore in the following section.
This can be partially mitigated by self-calibration or redundant calibration, although redundant calibration still suffers from this effect to some degree \citep{Byrne2019}.

We also make images of the pseudo-Stokes visibilities as a diagnostic tool.
The pseudo-Stokes visibilities \citep{Hamaker1996} are a linear sum of the linear polarization visibilities, defined as
\begin{align}
\label{eqn:stokes_vis}
\left(\begin{array}{c} V_I \\ V_Q \\ V_U \\ V_V \end{array} \right)
= \frac{1}{2} \left(\begin{array}{cccc} 1 & 0 & 0 & 1 \\
						       1 & 0 & 0 & -1 \\				
						       0 & 1 & 1 & 0 \\	
						       0 & -i & i & 0 \end{array}\right)
\left(\begin{array}{c} V_{\rm XX} \\ V_{\rm XY} \\ V_{\rm YX} \\ V_{\rm YY} \end{array} \right).
\end{align}
Note these are not true Stokes parameters, which are only properly defined in the image plane, but can be thought of as approximations to the true Stokes visibility one would form by Fourier transforming the true Stokes parameter from the image plane to the $uv$ plane.
In the limit that the instrumental (direction-dependent) Mueller matrix is the identity matrix, then the pseudo-Stokes visibility defined in \autoref{eqn:stokes_vis} is identical to the true Stokes visibility.
In practice, we do not expect this to be the case except for possibly near the pointing center in the image where, after having performed direction-independent calibration, we hope direction-dependent terms are minimal.

\begin{figure*}
\label{fig:gleam02_MFS_IQUV}
\centering
\includegraphics[scale=0.55]{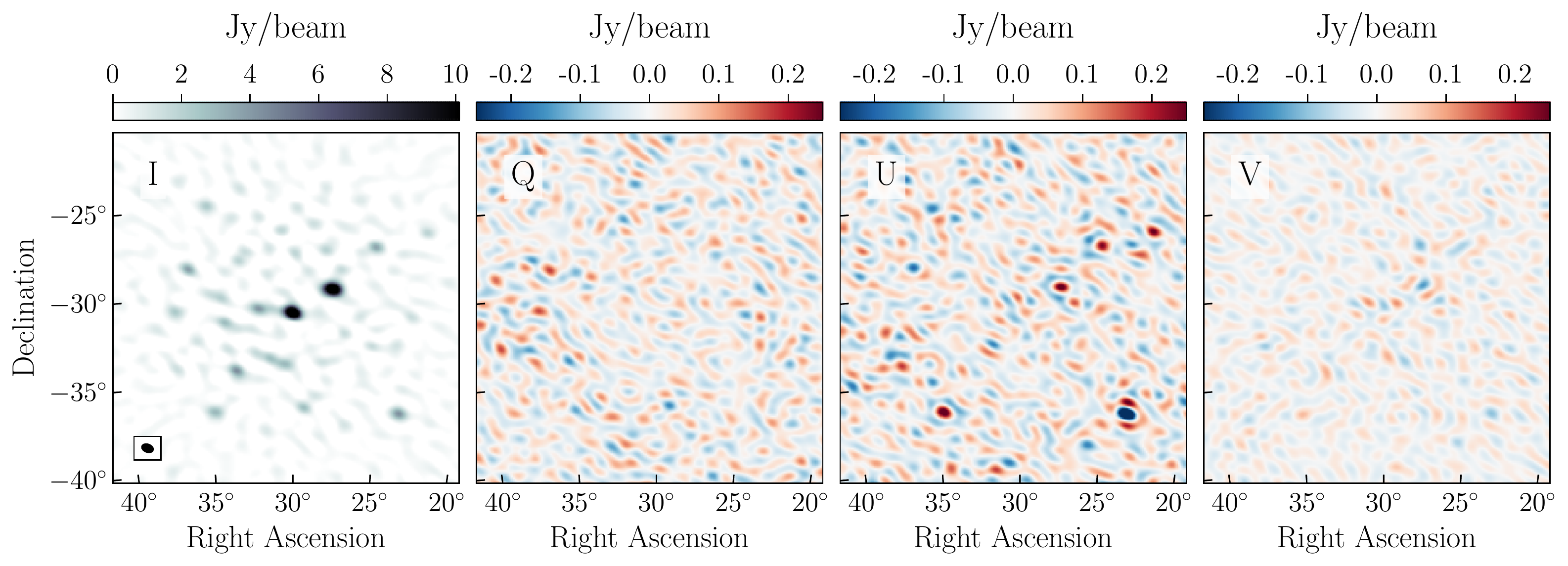}
\caption{Multi-frequency synthesis images (120 -- 180 MHz) of the GLEAM-02H field in all pseudo-Stokes I (far-left), Q (center-left), U (center-right) and V (far-right) polarizations.
Each image is CLEANed with the same parameters down to 1 Jy, with the CLEAN beam shown in the lower left.
Even with no polarization calibration, the observed leakage from I $\rightarrow$ Q, U \& V is a few percent.}
\end{figure*}

We do not expect appreciable levels of polarized sources in the GLEAM-02H field.
For a recent study by the MWA see \citet{Lenc2017}.
Given an ideal telescope with no instrumental leakage, we would therefore expect the pseudo Q, U and V visibilities to look noise-like.
However, we know that the primary beam response at a given point on the sky for the X and Y dipoles are not the same at low zenith angles, which will by itself cause polarization leakage of observed off-axis sources into Stokes Q \citep{Moore2017}.
Furthermore, we have not attempted to calibrate feed D-terms \citep{Hamaker1996} or the unconstrained relative X-Y phase parameter leftover after Stokes I calibration \citep{Sault1996}.
We also know from previous studies that mutual coupling exists at a non-negligible level \citep{Fagnoni2019, Kern2019b}, which is in principle a direction-dependent term in the Mueller matrix.
This means that we wholly expect that images formed from pseudo-Stokes visibilities will i) not necessarily be representative of the true Stokes parameters in the image plane, except for maybe near the pointing center and ii) that we should observe non-negligible amounts of polarization leakage from Stokes I $\rightarrow$ Stokes Q, U \& V.
To properly make true Stokes parameters one would image each of the linear dipole visibilities and perform direction-dependent corrections in the image plane before adding them in a similar manner as \autoref{eqn:stokes_vis}.
At the moment we defer this to future works that more carefully consider polarization calibration.

\begin{figure*}[t]
\label{fig:gleam02_PSF}
\centering
\includegraphics[scale=0.50]{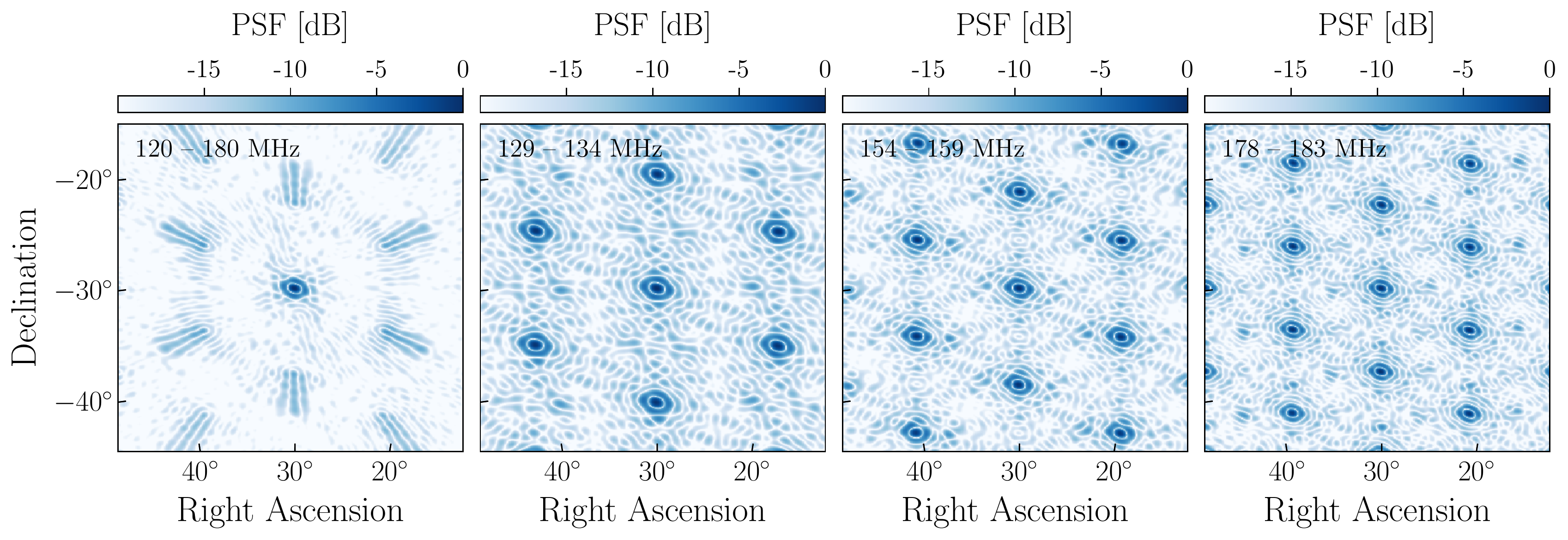}
\caption{The HERA Phase I point spread function (without primary beam correction) from a 5-minute observation across a wide band (left), and a narrow band located in a low-band spectral window (center-left), mid-band spectral window (center-right) and high-band spectral window (right). The grating lobes of the narrow band spectral windows appear in hexagonal patterns reflecting the (un-)sampled $uv$ spacings on the array, and reach upwards of 50\% of the peak PSF response at image-center.}
\end{figure*}

\autoref{fig:gleam02_MFS_IQUV} shows MFS images of the GLEAM-02H field in pseudo-Stokes I, Q, U \& V (left to right).
The first thing to note is that the observed leakage of Stokes I to Q, U and V is on the order of a few percent, which is quite low given we did not apply a polarization or direction-dependent calibration.
Looking at the pseudo-Stokes Q image we can see the effects of primary beam asymmetry between X and Y dipoles: without a primary beam correction (which is not applied here), the asymmetry will cause leakage of I $\rightarrow$ Q \citep{Moore2017}, which is exacerbated the more discrepant the primary beam responses are at a given point on the sky.
Although nearly azimuthally symmetric, the X-dipole beam is elongated along the North-South direction while the Y-dipole is elongated along the East-West direction \citep{Fagnoni2019, Martinot2018}.
This means we might expect the relative amplitude of the X and Y beams to attain a better match in the corner of our images, and would therefore expect to see more I $\rightarrow$ Q leakage in a quadrupolar pattern on the sky.
Indeed, this is observed in the pseudo-Stokes Q image to some degree (\autoref{fig:gleam02_MFS_IQUV}).

The pseudo-Stokes U and V images also exhibit interesting behaviors, in particular the sources in the pseudo-Stokes U image that are clearly correlated with true Stokes I sources, as well as the rumble in the pseudo-Stokes V image that seems to be concentrated near the main lobe.
This could be due to polarization leakage stemming from the uncalibrated X-Y phase term, however, further work is needed to identify its exact cause.

Having shown that our calibration does a fairly good job bringing our data in-line with our model (\autoref{fig:gleam02_MFS_XX}) and that, even without polarization calibration, polarization leakage is observed at a few percent (\autoref{fig:gleam02_MFS_IQUV}), we should also show that our derived bandpass is an accurate solution as a function of frequency.
To do this we can make a spectral cube of our calibrated data and compare to the original catalogue used for calibration.
However, making a spectral cube with fine frequency resolution means that the point-spread function (PSF) sidelobes and grating lobes become increasingly a problem due to the sparse sampling of the $uv$ plane.
\autoref{fig:gleam02_PSF} shows the HERA Phase I PSF across a wide 60 MHz band (left) and three narrower 5 MHz bands (center and right).
For wide-band imaging the PSF grating lobes are smeared out due to the large bandwidth.
However, for narrow-band imaging the grating lobes rise to above 50\% the peak PSF response at image center; for narrower spectral windows this is only exacerbated.
Such strong grating lobes make performing deconvolution to high dynamic range difficult, especially in a confusion-limited regime.

\begin{figure*}
\label{fig:gleam02_pI_spec}
\centering
\includegraphics[scale=0.60]{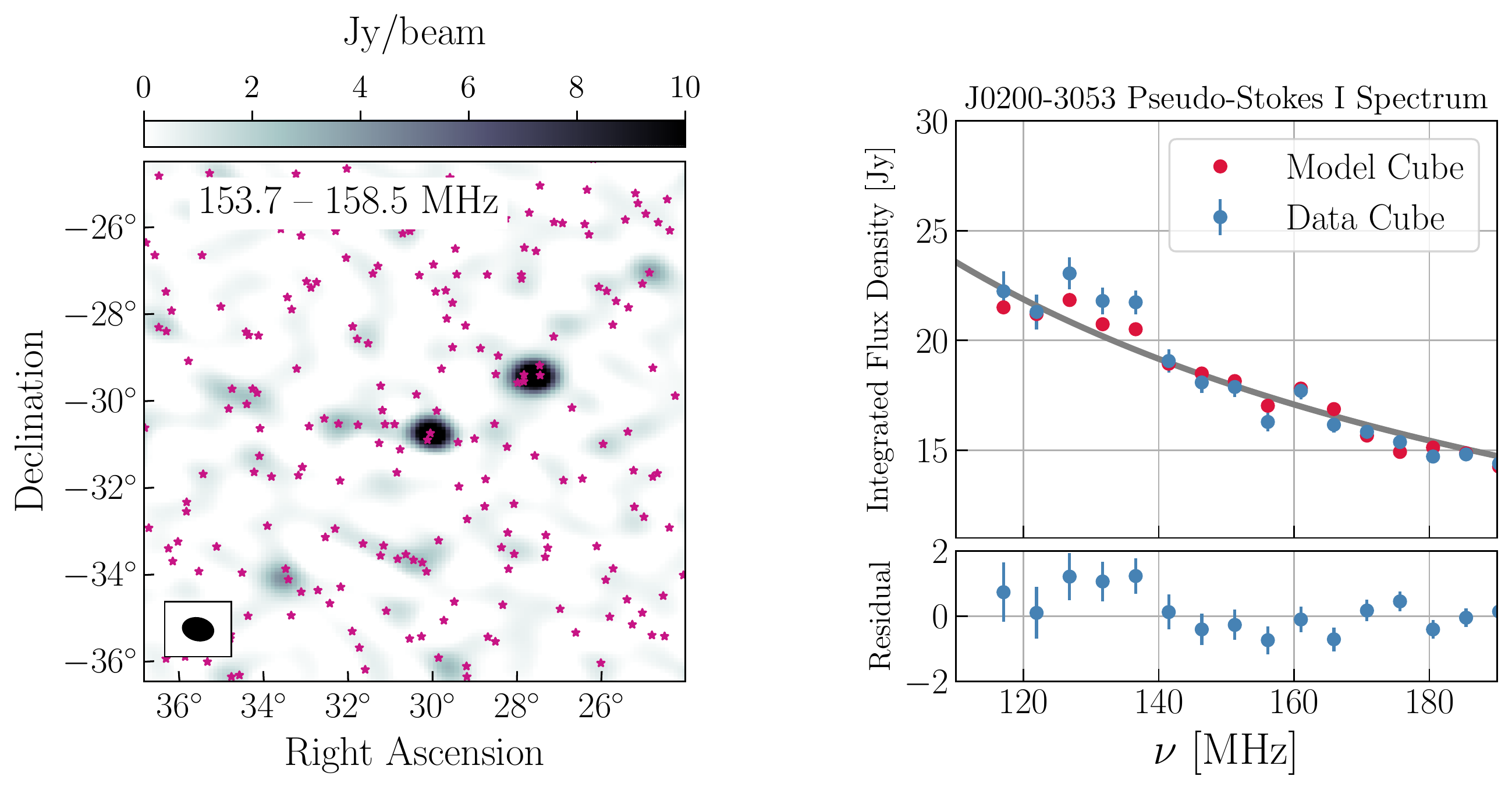}
\caption{The extracted spectrum of the primary calibrator GLEAM J0200-3053 from the GLEAM-02H field.
{\bfseries Left}: CLEANed MFS image of the data (colorscale) across a narrow band (153.7--158.5 MHz). The purple markers show each GLEAM point source above 0.5 Jy used in the initial model, demonstrating the degree of source confusion given the Phase I angular resolution.
{\bfseries Right}: Extracted spectrum of J0200-3053 (center of left image) across each channel in the data spectral cube (blue) and model spectral cube (red). The data and model are in good agreement with each other, and are well-fit by the original input GLEAM J0200-3053 power law model (grey).
Large-scale frequency deviations from the power-law fit are partially reflected in both the data and model, suggesting that they are not due to mis-calibration but due to imperfect PSF sidelobe removal in the process of imaging.
The data cube -- model cube difference shows residual structure at the $\sim5\%$ level.
}
\end{figure*}

We can partially work around this by applying CLEAN masks around bright sources and then CLEANing down iteratively while opening up the mask to dimmer and dimmer sources.
Indeed, this is what we do to make a coarse-channel spectral cube, which consists of MFS images with 5 MHz in bandwidth using iterative CLEAN runs targeting successively dimmer sources.
However, in the case of single-channel imaging even this does not work: the grating lobes are just too severe to deconvolve them from the image without misplacing source flux in un-modeled sidelobes.
\autoref{fig:gleam02_pI_spec} shows the result of a coarse, 5 MHz wide spectral cube CLEAN for a spectral window centered at 155 MHz (greyscale, left).
We also show all GLEAM sources in the original model with fluxes above 0.5 Jy in purple, which demonstrates the high degree of confusion given our modest angular resolution: each ``source'' in our images are generally two or more GLEAM sources blended together.
We therefore cannot easily relate the source flux in our images to one or even multiple sources in the GLEAM catalogue, as each GLEAM source will have a different contribution to a HERA source given its distance from it and the HERA PSF.
If our goal is to compare extracted fluxes between the data and a point-source model we should take the PSF out of the equation.
The deconvolution on the data attempts to do this at some level, but is limited fundamentally in precision by the width of the synthesized beam.
Another way is to simply add the PSF into the model by imaging the model visibilities and then CLEANing and running a source extraction in the same way as is done for the data.
This means that the inherent shortcomings of the deconvolution and the limitations of the PSF (both things not really relevant for validating gain calibration done in the $uv$ domain rather than the image domain) are kept constant between data and model, so we can make a better comparison between the two.

Source extraction is done on a source-by-source basis with custom software.
First we select the coordinates of a desired source in the data, then the extraction process makes a postage-stamp cutout in the shape of the synthesized beam with twice its FWHM around the desired source and fits a 2D Gaussian of variable major axis length, eccentricity, amplitude and position angle using the \texttt{astropy.modeling} module.
It then records the integrated flux of the fit in Jy and computes the fit error by taking the RMS of the image in an annulus outside the cutout and dividing by the square-root of the synthesized beam area \citep{Condon1997}.

\begin{figure*}[t!]
\label{fig:data_model_wedge}
\centering
\includegraphics[scale=0.48]{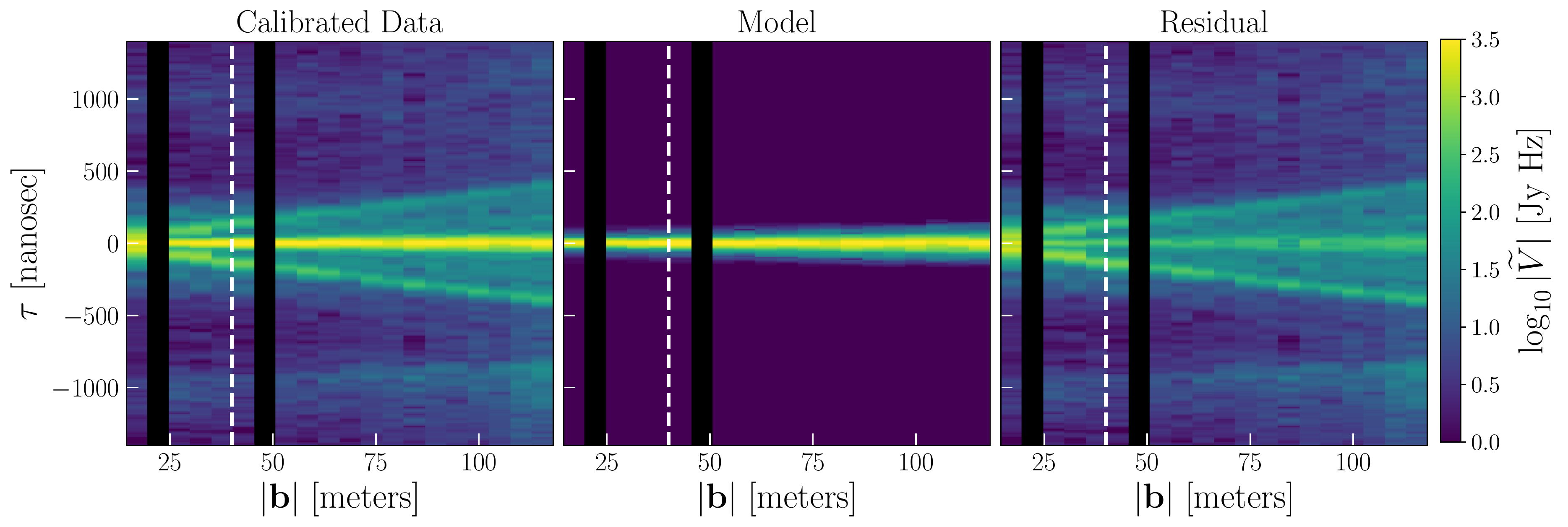}
\caption{Redundantly-averaged pseudo-Stokes I visibilities in delay space transformed across a 120 -- 180 MHz spectral window, ordered according to baseline length.
We show the calibrated data (left), the point source calibration model (center), and their residual (right). Short baselines to the left of the white dashed line are not used in calibration. Black regions represent a lack of data at those baseline lengths. The data clearly show a pitchfork-like foreground wedge predicted by \citet{Thyagarajan2016}. Note that the edges of the pitchfork are not reflected in the calibration model, which will generate calibration errors. The residual power of the main foreground lobe in the wedge is suppressed by about a factor of 10 compared to the data, but is still seen above the noise floor of the data. Additional power at large delays ($\tau\sim1000$ ns) are the same systematics seen in \citet{Kern2019b}.}
\end{figure*}

This is done for the GLEAM-02H field primary calibrator J0200-3053 for each 5 MHz-wide channel in the coarse spectral cube of the data and model, shown in \autoref{fig:gleam02_pI_spec}.
The data (blue) and model (red) are in good agreement with each other across the entirety of the band, and are in relatively good agreement with the primary calibrator's original power law model from the GLEAM catalogue (grey, \autoref{tab:calibrators}).
Both the data and model exhibit sinusoidal frequency fluctuations about the power law model; however, because this structure is represented in model spectra we can conclude that some of these fluctuations are due to imperfect PSF sidelobe removal in the CLEAN process, rather than calibration errors.
If we take the difference between the extracted data and model fluxes then we see residual deviations at $\sim5\%$ the source's intrinsic flux.
However, these deviations look similar in form to the first-order sinusoidal variations about the smooth power law, possibly suggesting that some of these features in the residual are too due to imperfect PSF sidelobe removal.
One possibility is that the CLEAN deconvolution achieved better sidelobe removal on the model cube compared to the data cube, which would generate the kind of observed sinusoidal variations in the data-to-model residual.
This would not be entirely surprising given the extra terms in the data that are not in the model, including diffuse foregrounds, which would make deconvolution more difficult.
The second-order fluctuations in the data-to-model residual (channel-to-channel) hovers at roughly 1\% of the intrinsic source flux.
Overall, these lines of evidence suggest that the quality of the spectral calibration across the band is on the order of a few percent.

However, the leading uncertainty in our absolute calibration is the determination of the overall flux scale.
By adopting the GLEAM point source catalogue as our model, we have set the flux scale of our calibration to GLEAM, which themselves tie their flux scale to the VLA Low-Frequency Sky Survey redux \citep[VLSSr;][]{Lane2014}, the NRAO VLA Sky Survey \citep[NVSS;][]{Condon1998} and the Molongo Reference Catalogue \citep[MRC;][]{Large1981}.
When comparing their measured source fluxes to sources from these catalogues, their flux scaling appears to be unbiased with an uncertainty of $\sim10\%$.
One concern about our usage of a single GLEAM field to set the flux scale is the fact that GLEAM's J0200-3053 source may be an outlier in that distribution, implying that our flux scale could be significantly biased.
This concern is tempered by the residual image of \autoref{fig:gleam02_MFS_XX}, which shows that not only J0200-3053, but all sources in the main lobe of the beam have an unbiased residual, meaning that our final flux scale agrees with the GLEAM flux scale for all sources in the main-lobe of our primary beam.

To better understand the match between the data and the flux density model, we take the full gain solutions from the GLEAM-02H field and use them to calibrate all baselines in the data.
We then form pseudo-Stokes I visibilities and coherently average all baselines within a redundant group (i.e. with the same baseline length and orientation).
Then we take the Fourier transform of the visibilities across a wide bandwidth spanning 120 -- 180 MHz, having first applied a Blackman-Harris windowing function \citep{Blackman1958} to limit spectral leakage in the discrete Fourier transform (DFT).
Before we do this, however, we must first account for the frequency channels that have been flagged due to RFI.
These will create strong sidelobes in the Fourier transform if not accounted for.
To overcome this, we employ a 1-dimensional deconvolution algorithm that deconvolves the sidelobes due to the RFI, which is conceptually identical to the CLEAN algorithm employed by radio interferometric imaging to interpolate over missing $uv$ samples \citep{Hogbom1974}, and can be found in the \texttt{hera\_cal} package.
In our case, we build a CLEAN model in delay space out to $\tau = 2000$ ns, and interpolate over the flagged channels with the CLEAN model before taking the final DFT to the delay domain, which we do for the data and model visibilities in an identical fashion.
We then coherently time average the 5 minutes of data around the GLEAM-02H calibration field, take the absolute value of the averaged visibilities and average all baselines of the same length, regardless of orientation.
This is the same procedure one would take to form 2D cylindrically averaged power spectra, but in this case we are working with just the visibilities in the Fourier domain.

\autoref{fig:data_model_wedge} shows this for the calibrated data (left), model (center) and shows their residual (right).
From it, we can clearly see the pitchfork-like foreground wedge with a main component centered at $\tau=0$ ns and then branches at positive and negative delay following the horizon line of the array, which is not plotted for visual clarity and is explained in more detail in \autoref{sec:pspec}.
Recall that baselines shorter than 40-meters in length (white dashed) are not used in calibration.
The pitchfork branches are caused by the foreshortening of a baseline's separation vector at the horizon, thus increasing its sensitivity to diffuse emission \citep{Thyagarajan2015a}.
The point source model, lacking diffuse foregrounds, clearly does not have a strong pitchfork feature.
This discrepancy will create gain errors in the calibration solutions at the delay scale of the pitchfork, which for baselines above 40-meters begins at around 150 ns and extends beyond that for longer baselines.
This is explored in the following section.
Lastly, the data and model are somewhat well matched at $\tau\sim0$ ns, with the residual power being suppressed by a factor of 10 compared to the data but still above the noise floor of the data outside the wedge.
This residual power can come from un-modeled diffuse flux in the main lobe of the primary beam, but is also likely to be from calibration errors due to mis-modeled point sources.
An increase in observed power in the data at large delays $|\tau| > 800$ ns is a cross coupling systematic, and is not foreground signal \citep{Kern2019b}.

\begin{figure*}
\label{fig:gleam02_raw_gains}
\centering
\includegraphics[scale=0.52]{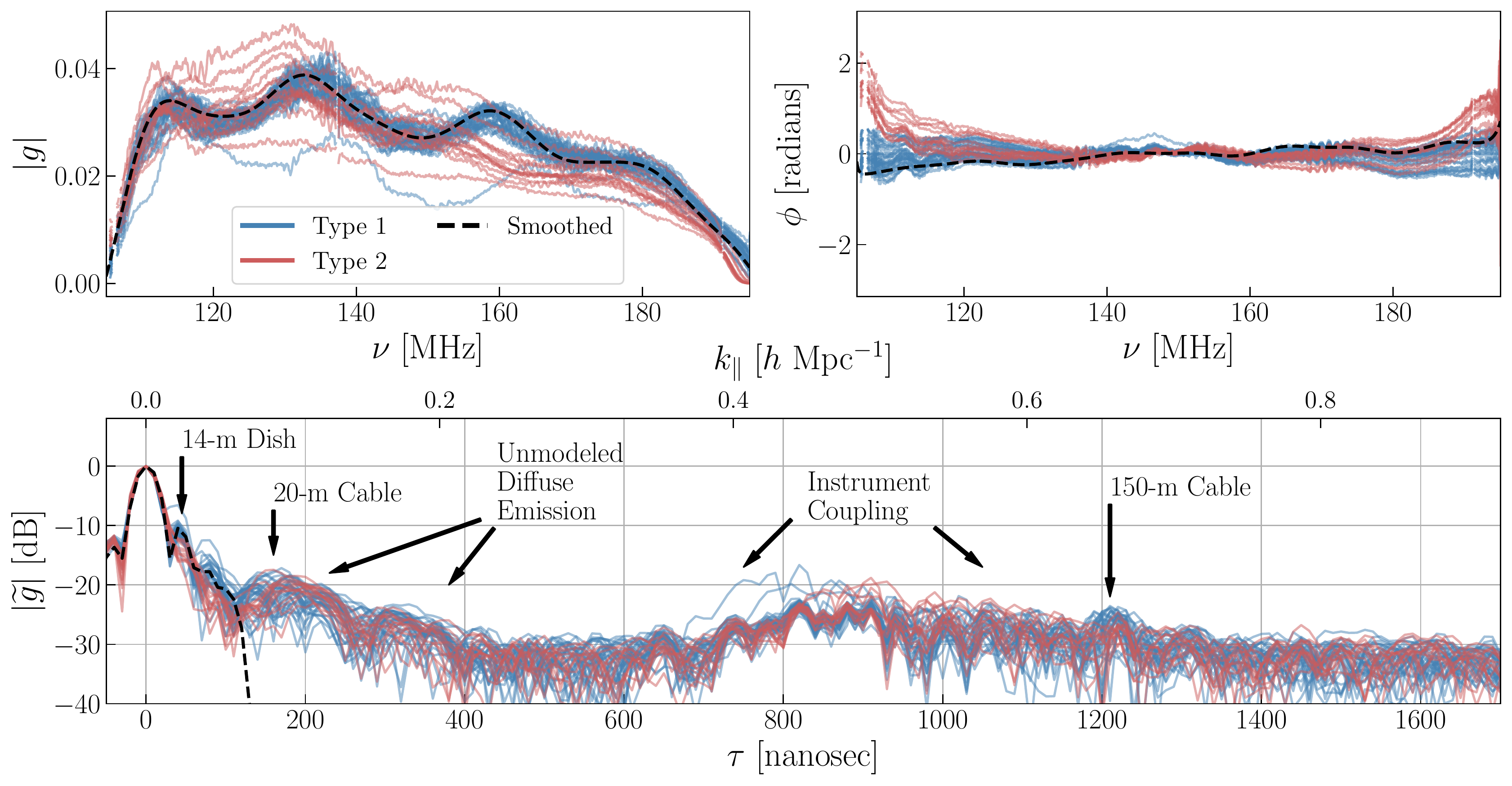}
\caption{Antenna gains derived from the GLEAM-02H field. Type 1 \& 2 signal chains are plotted in blue and red, respectively. The phase of the gains (top-right) are plotted after taking out the cable delay from each antenna for visual clarity.
The peak-normalized delay response of the gains show structure at delays representative of elements in the signal chain (bottom), and also show contamination by terms that are not antenna based, like un-modeled diffuse emission and instrumental coupling systematics \citep{Kern2019b}.
We also show one of the Type 1 gains smoothed at a 100 nanosecond scale for reference (dashed-black).}
\end{figure*}

%%%%%%%%%%%%%%%%%%%%%%%%%%%%
%%%%%%%%%%%% Gain Stability %%%%%%%%%
%%%%%%%%%%%%%%%%%%%%%%%%%%%%
\section{Gain Stability}
\label{sec:gains}
In this section we characterize the spectral and temporal properties of the derived complex antenna gains and discuss their impact on downstream analyses.
Gain calibration is a multiplicative term in the frequency and time domain meaning it can equivalently be thought of as convolution in the Fourier domains of delay and fringe-rate, the Fourier duals of frequency and time respectively, by a ``gain kernel,'' or the Fourier transform of the gain response.
Solving for and applying antenna-based gains can therefore be thought of as trying to deconvolve the inherent gain kernel imparted by the instrument.
For \tocm experiments aiming to uncover a signal buried under noise and systematics, the principal concern when applying gain solutions to the data is understanding how this gain kernel may or may not be smearing foreground signal to spectral modes that are otherwise foreground-free:
any kind of deviation in the derived gain solution from the true underlying gain will cause such smearing, at some level.

\subsection{Spectral Response}
\label{sec:freq_stability}
Works investigating sky-based calibration in the limit of an incomplete sky model showed it results in gains with erroneous spectral structure that can fundamentally limit \tocm studies \citep{Barry2016, Ewall-Wice2017, Byrne2019}.
Similar effects have been shown to exist for redundant calibration, where inherent non-redundancies of the array create a similar type of spectrally-dependent gain error \citep{Orosz2019}.
What has yet to be studied in detail is how other kinds of instrumental systematics, such as mutual coupling or crosstalk, get picked up in the process of gain calibration and what their effect is in shaping the inherent and estimated gain kernel.
For systematics like crosstalk and mutual coupling which are highly baseline-dependent, one would naively expect that the antenna-based gains would not significantly pick up on these terms due to their decoherence when averaged across different baselines; however, it would not be surprising to see them at some level reflected in the gain solutions, even if they are averaged down to some degree.
Furthermore, \autoref{fig:data_model_wedge} shows us that there is a non-negligible data-to-model discrepancy caused by un-modeled diffuse emission even for baselines above our 40-meter cut, which will also create gain errors.

To summarize \citet{Kern2019b}, the HERA Phase I system shows evidence for cross-coupling systematics at large delays $|\tau| > 800$ ns, and also shows evidence for diffuse flux and / or mutual coupling at smaller delays corresponding to a baseline's geometric horizon (for $|\mathbf{b}|=45$ m, this is $\sim150$ ns).
In \autoref{fig:gleam02_raw_gains} we show the frequency and delay response of the CASA-derived, sky-based gains from \autoref{sec:skycal}.
We plot the gain amplitude (upper-left), gain phase after removing the cable delay for each antenna (upper-right) and the Fourier transform of the gains in delay space (bottom) normalized to their peak power at $\tau=0$ ns.
We categorize the gains into Type 1 (blue) and Type 2 (red) signal chains (\autoref{fig:uvplot}), which shows a clear bi-modality in the spectral structure of the gains between these groups.
This bi-modality is also seen in the reflection properties of the signal chains and is discussed in more detail in \citet{Kern2019b}.
Arrows mark the expected regions in delay space where certain electromagnetic elements in the signal chain can create systematics, such as reflections within the 14-meter dish and reflections in the 20-meter and 150-meter coaxial cables.
The gain kernels of each antenna (\autoref{fig:gleam02_raw_gains}-bottom) also clearly shows that instrumental cross coupling systematics at $|\tau| > 800$ ns are being picked up by the gain solutions.
It also shows un-modeled diffuse emission at lower delays $|\tau| > 200$ ns, which may also have contributions from mutual coupling systematics that appear at similar delays.
Because instrumental cross coupling and diffuse emission are baseline-based and not antenna-based, they cannot be calibrated out of the data with antenna-based, direction-independent gains, and must be removed on a per-baseline basis.
This means that the presence of these structures in the gains will only spread the systematics around: in the worst case spreading them to baselines that may have been systematic free to begin with.

\begin{figure*}
\label{fig:gleam02_fullcal_dly_frate}
\centering
\includegraphics[scale=0.70]{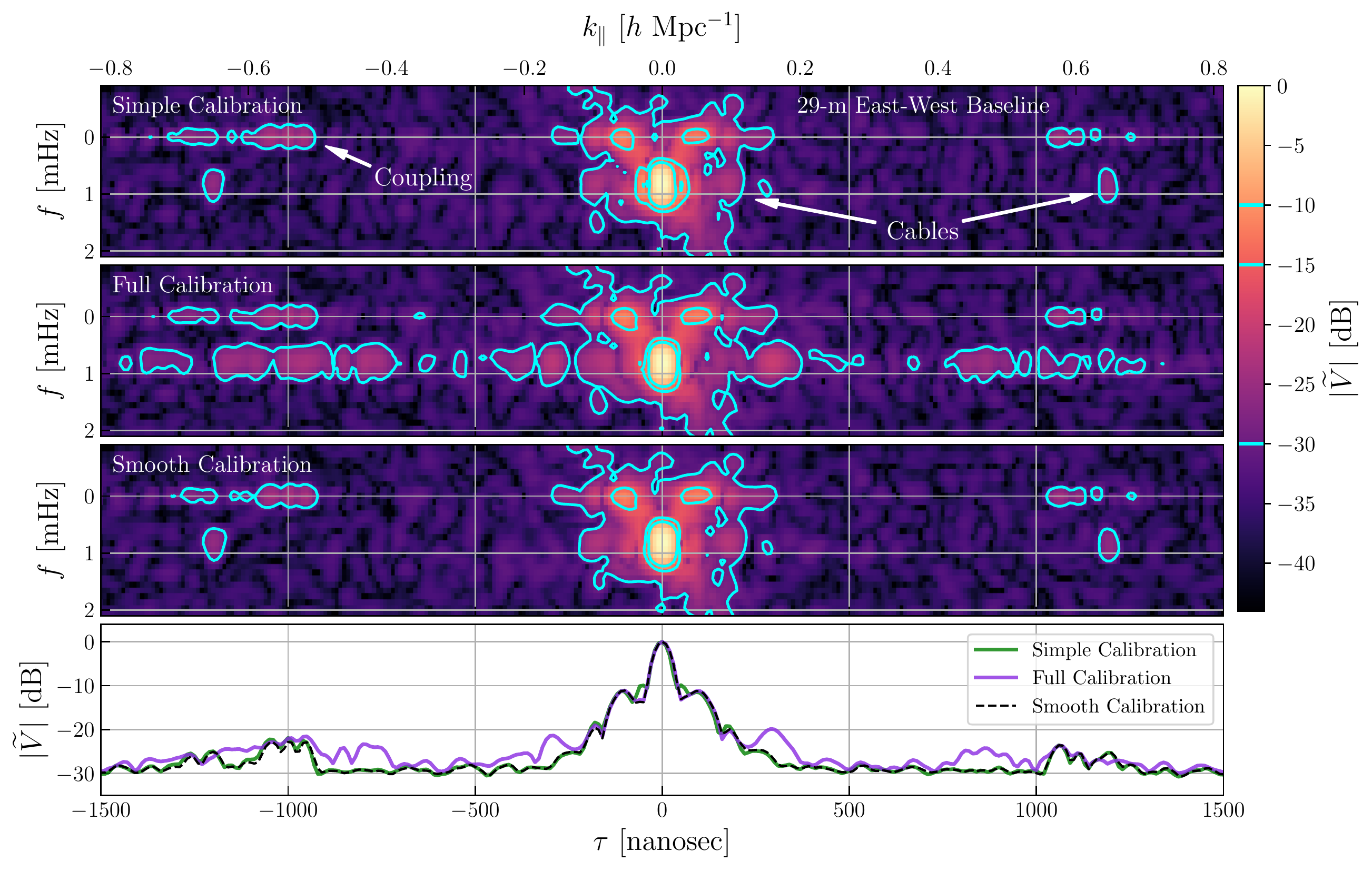}
\caption{Sky-based gains applied to a single 29-meter East-West visibility over 8-hours of LST and transformed to delay and fringe-rate space (see text for details). The data are peak-normalized, and contours show -30, -15, and -10 dB levels. The time-averaged delay responses are shown in the bottom panel. Sources of cable reflection and instrument coupling are marked. The full gain applied to the data leads to significant contamination of coupling systematics due to the full gain kernel smearing the foreground horizontally in fringe-rate and delay space. Smoothing the calibration allows us to calibrate out the features at low delays we know to be calibrate-able (e.g. dish reflections) and toss out features in the gain kernel above 100 nanoseconds.}
\end{figure*}

\autoref{fig:gleam02_fullcal_dly_frate} shows the result of applying sky-based gains to the visibility data and transforming to the Fourier domains of delay and fringe-rate space.
We apply the gains to 8-hours of drift-scan data from a single 29-meter East-West visibility, and we do this having filtered the gains in three different ways: 1) the first method (simple calibration) takes only the band-averaged amplitude and cable delay component of the gain 2) the second method (full calibration) just takes the full gain solution as-is, and 3) the third method (smooth calibration) smooths the gains across frequency out to a 100 ns scale, which is also plotted in \autoref{fig:gleam02_raw_gains} (black-dashed).
The bottom panel shows the time-averaged delay response of the panels shown above.
In the simple calibrated data, the foregrounds are contained to low delays and appear predominately at positive fringe-rates, which we expect because the sky rotates in a single coherent direction in the main lobe of primary beam \citep{Parsons2016}.
Foregrounds can also occupy near-zero and negative fringe-rate modes, which correspond to structures on the sky near the South celestial pole and near the horizon, but are attenuated by the primary beam response.
If the data were nominal then the rest of the Fourier space would be dominated by thermal noise; however, this is not what we observe.
We also see cable reflection signatures, which should appear as reflected copies of the foregrounds at the same fringe-rates but at positive and negative delays (marked).
And we see strong cross coupling features at large positive and negative delays occupying near-zero fringe-rate modes (marked).

\begin{figure*}
\label{fig:smooth_full_resid}
\centering
\includegraphics[scale=0.48]{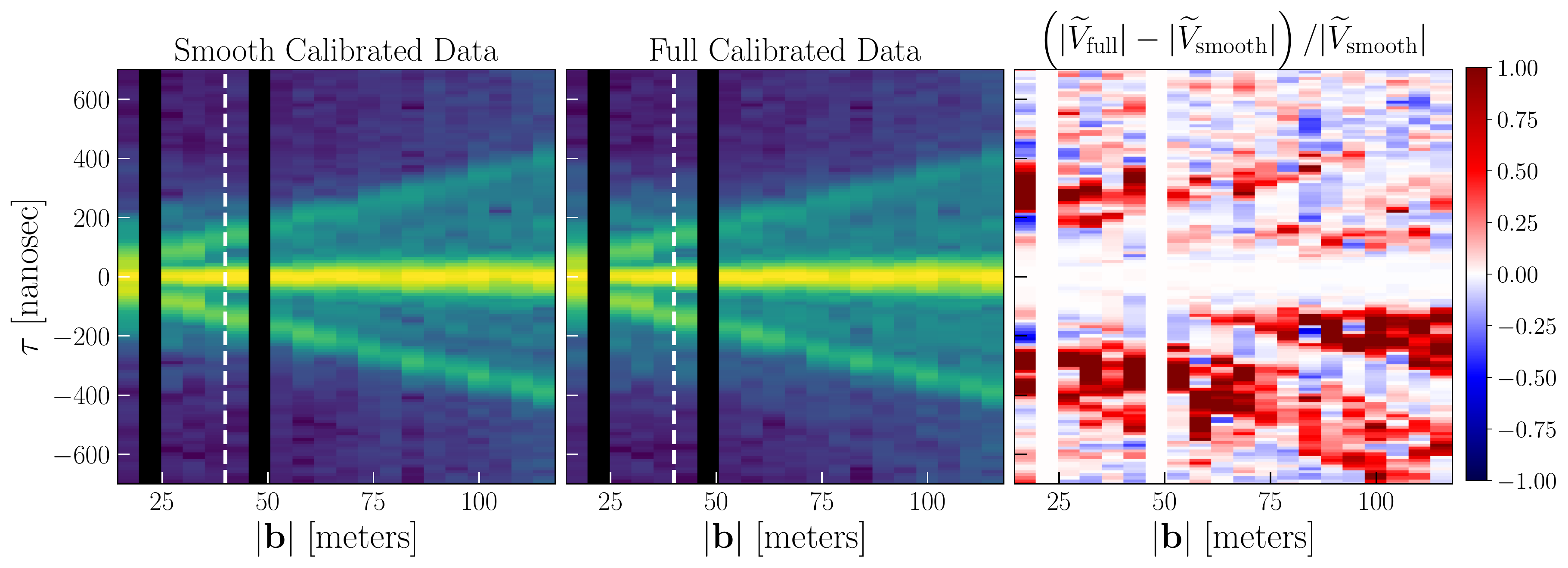}
\caption{Redundantly-averaged pseudo-Stokes I visibilities in delay space transformed across a 120 -- 180 MHz spectral window, ordered according to baseline length.
We show the smooth calibrated data (left), full calibrated data (center) and their fractional residual. The calibrated data (left and center) are plotted on the same colorscale as \autoref{fig:data_model_wedge} over a smaller delay range to highlight the features within the foreground wedge. Within the smoothing scale of 100 ns, the fractional residual shows the two are in good agreement as expected. Outside the smoothing scale, however, the residual shows significant excess structure (red) in the full calibrated data not seen in the smooth calibrated data, which suggests that the structures are not real and are errors in the gain solution.}
\end{figure*}

When we go to apply the full calibration we find a large amount of excess structure at intermediate and large delays occupying positive fringe-rates, which is not surprising given the gain kernels shown in \autoref{fig:gleam02_raw_gains}.
We see that other baselines that happened to have the systematics at intermediate delays have contaminated this baseline at the same delays.
What is more, these systematics are now occupying the same positive fringe-rate modes as the sky,\footnote{This happens because the gains are a multiplicative term, meaning that although the systematics originally occupied $f\sim0$ Hz, the contaminated gains spread them to $f > 0$ Hz modes.} and therefore cannot be easily removed with standard cross coupling removal techniques \citep{Kern2019a}.
There are some benefits of the full calibration, though.
One is that it can calibrate out signal chain reflections because those factor as antenna based terms.
This can be seen in the data as the suppression of the cable reflections at large positive and negative delays, and also in the suppression of the dish reflection at $\tau=\pm50$ ns (which is most apparent as the tightening up of the contours in the brightest spots of the foregrounds, or the drop in the shoulder power in the time-averaged spectra).
While cable reflections at high delays can be calibrated out with sky-agnostic modeling \citep{Ewall-Wice2016b, Kern2019a}, calibrating out reflections at low delays that bleed into the main foreground lobe is harder, and thus better suited to correction via standard gain calibration.

The ideal compromise, then, is to smooth our gains to keep the gain kernel at low delays and suppress its power at delays that we no longer trust its response.
For the calibration at hand, this seems to be at roughly 100 nanoseconds, which enables the calibration to pickup on the dish reflection at 50 ns but suppresses the spurious terms in the gains at 150 ns and beyond.
Given our 100 MHz bandwidth with 1024 channelization, a maximum delay range of 100 ns leads to about 15 free delay modes in the smoothed gains, which can be thought of as a smoothed gain with 15 spectral degrees of freedom per antenna and dipole polarization.
Applying this gain to the data (last two panels in \autoref{fig:gleam02_fullcal_dly_frate}), we see that we recover the best of both scenarios: the dish reflection is suppressed as desired and we also do not spread more instrument coupling at intermediate and high delays over what is already present in the data.
To perform the smoothing we use the same delay-domain deconvolution technique described before as a low-pass Fourier filter, which is useful given that the gains are also flagged at certain frequency channels due to RFI.
Although this calibration is performed for a single time, one can also take time and frequency dependent calibration solutions and smooth across both the temporal and spectral axes with this technique.

We can also show the effects of the smooth and full calibration on the full dataset.
We do this by applying the calibration to the data and transforming them to the delay-domain in a similar manner as was done for \autoref{fig:data_model_wedge}.
In this case, \autoref{fig:smooth_full_resid} plots this for the smooth calibrated data (left), the full calibrated data (center) and their fractional residual (right).
Note that the calibrated data are plotted on the same colorscale as \autoref{fig:data_model_wedge}, but are plotted with a smaller delay range to highlight features within the foreground wedge.
We see that the two calibrations achieve a good match at low delays, as expected, but for delays beyond the smoothing scale we find that the full calibrated data has significant excess structure (red) compared to the smooth calibrated data.
This is indicative of the full calibration \emph{introducing} spectral features into the data, rather than calibrating them out, which is highly suggestive of gain errors on these scales and further motivates the $\tau\sim100$ ns smoothing scale of the gains derived in \autoref{sec:skycal}.

Philosophically, this kind of approach to gain calibration, in other words keeping only degrees of freedom like low delay modes that we trust and filtering out the rest, is conservative from the perspective of not introducing structure into the data that was not already there.
The cost of this approach is that we are not calibrating out gain structure at these delays inherently introduced by the instrument, if it exists in the first place.
At the moment, however, we do not really have much of a choice: providing a constrained calibration with a few degrees of freedom is the best we can currently do, and until we have evidence that structure in the gain kernel at higher delays is real gain structure, we should not attempt to calibrate it out.
This approach makes interpreting a fiducial detection in the power spectrum at similar, intermediate delays somewhat convoluted, and a suite of null tests and jacknives will be necessary to try to tease out whether said detection is residual calibration structure or real sky structure.
%HERA is not unique is taking this approach: analyses from the MWA and LOFAR show that gain smoothing is needed to restrain the derived gain kernel from spreading foreground power into the EoR window.

The obvious question moving forward for HERA then is, do we believe there to be true gain structure at low and intermediate delays that we need to calibrate out?
The answer to this depends on the required dynamic range.
For low delays we generally need $10^5$ in dynamic range performance of the gain kernel due to the foreground-to-EoR amplitude ratio: for larger delays this requirement becomes more stringent as the EoR signal is expected to weaken.
Therefore, do we think HERA has true gain structure at some level above -50 dB at $\tau\sim$200 ns?
Based on simulations \citep{Fagnoni2019} and a rough extrapolation of \autoref{fig:gleam02_raw_gains} the answer is probably yes, and therefore, we need a way to remove the cross coupling systematics from the data \emph{before} performing antenna gain calibration.
Cross coupling systematic removal is done by applying a high-pass filter in fringe-rate space \citep{Kern2019a, Kolopanis2019}.
This removes cross coupling, which occupies low fringe rates, but it also removes a component of the foregrounds as well, which we need for calibration.
Doing this only on the data and not on the model would create a discrepancy in the data that would act as its own form of systematic.
Fringe rate filtering therefore needs to be done on the model and data before calibration in order to probe the true instrument gain kernel to higher and higher delays.
Achieving high fringe-rate resolution for a high-pass filter means simulating a large LST coverage with a wide-field flux density model.
Unfortunately, the CASA-based calibration methodology presented in this work does not easily lend itself to this as it only reliably simulates short time intervals near the calibration field.
This kind of analysis is best done using a numerical visibility simulator with wide-field diffuse and point-source maps, which we defer to future work.

Other smoothing algorithms have been investigated in the literature, which has been motivated due to a recent understanding of how incomplete sky models cause gain errors in sky-based calibration \citep{Barry2016, Byrne2019} and non-redundancies cause gain errors in redundant calibration \citep{Ewall-Wice2017, Orosz2019}.
The MWA, for example, uses low-order polynomials to smooth their sky-based gain solutions to limit gain error spectral structure in 21\,cm power spectral analyses \citep{Beardsley2016, Barry2019a}.
The reason we opt for direct Fourier filtering of the gains in this work is because a truncated polynomial basis is not able to encapsulate arbitrary gain fluctuations on large scales; in other words they do not form a complete basis in the Fourier domain for low-delay modes.
This is fine for mitigating small-scale structure but means one runs the risk of not calibrating out large-scale modes that can cause biases in narrow band power spectrum analyses, although in simulated MWA analyses there is no evidence for such biases \citep{Barry2016}.

\begin{figure*}
\label{fig:temp_oscillations}
\centering
\includegraphics[scale=0.50]{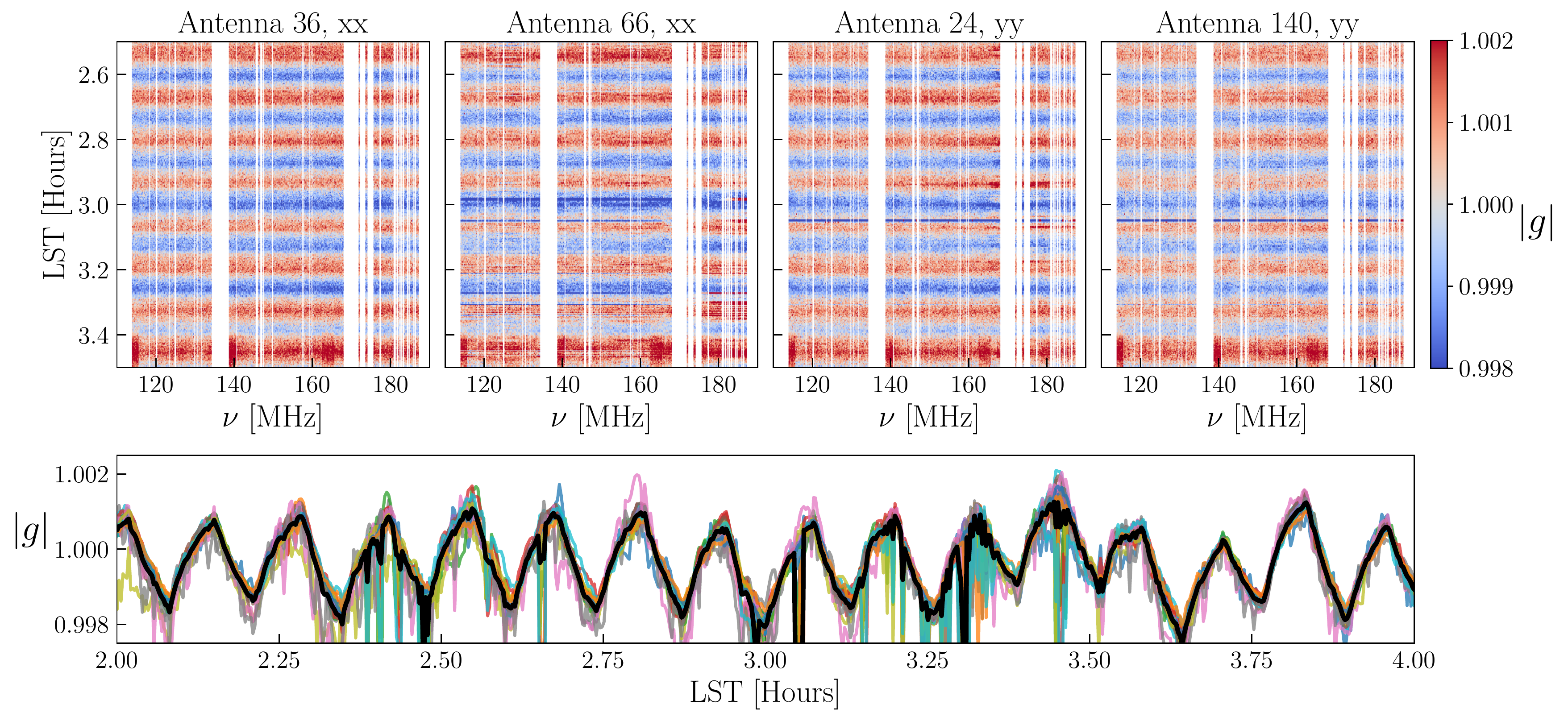}
\caption{Temperature oscillations in the instrumental gain due to an air-conditioning cycle in the field container housing the ADC are a 0.1\% effect. Top panels show the square-root of the ratio of the raw auto-correlations to the time-smoothed auto-correlations for a few antennas and both XX and YY polarization. The oscillation looks to be of roughly the same amplitude across different antennas, polarizations and frequencies.
The bottom panel shows the frequency-averaged oscillation for a handful of antennas (colored lines) and their average (black). This shows a saw-tooth time profile that also matches temperature data collected in the container.}
\end{figure*}

\subsection{Temporal Response}
\label{sec:temporal_response}
In this section we use the data to assess the temporal stability of the instrumental gain.
HERA observations are taken in drift-scan mode, meaning the array does not change or move over the course of observations.
This lends itself to a fairly stable instrument as a function of time, and we therefore do not expect large deviations in the gains over short time intervals.
However, effects such as ambient temperature drift and the cooling cycle within signal chain nodes are known to cause slight drifts in the calibration over the course of a night \citep[e.g.][]{Jacobs2013}.
In this section we investigate the data to quantify the amplitude of these gain drift terms and confirm they can be calibrated out if necessary.
Note we do not actually apply time-dependent gains to the data in the remainder of this work: we merely present ways in which these terms can be calibrated-out for deep integrations if necessary.

All signal chains in the HERA array are brought via coaxial cable to an RFI-shielded and air-conditioned container in the field where the data are converted from analog to digital signals and are then correlated.
Due to the air-conditioning cycle within this container, which cycles at roughly a 6-minute period, we expect the overall amplitude of the gains to drift at the same timescales.
We can estimate the amplitude of this drift using a smoothed version of the auto-correlations, which is the approach adopted by the LWA \citep{Eastwood2019} which faces the same issue.
Assuming that the only temporal structure in the auto-correlations occurs intrinsically at the time-scale of the beam crossing time ($\sim$40 minutes), we can probe time structure from the gains by taking a time-smoothed version of the auto-correlation and dividing it by the un-smoothed auto-correlation.
We smooth a handful of auto-correlation visibilities on a 20-minute timescale, divide their un-smoothed visibility counterparts by them and take their square-root, which leaves us with a set of ratio waterfalls as a function of frequency and time for each antenna-polarization.
We show some of these in \autoref{fig:temp_oscillations}, which plots the square-root ratio for each time and frequency bin for four antennas (top row) and also their frequency-average as a function of LST (bottom panel).
We see that the gain fluctuations induced by the air-conditioning cycle in the container has a coherent phase and amplitude across all antennas and polarizations, and is also fairly constant across frequency.
The frequency-average of each antenna and their respective average is shown to reflect a sawtooth profile as a function of time, whose profile inversely matches temperature data collected within the container.
\autoref{fig:temp_oscillations} shows us that the 6-minute gain oscillations are a very small effect at the 0.1\% level, and can be decently well-calibrated by a single number as a function of time for all antennas, polarizations and frequency channels in the array.
The HERA Phase II configuration will have a forced air cycling system that will better control fast temperature variations in container units.

\begin{figure}
\label{fig:avg_gain_drift}
\centering
\includegraphics[scale=0.50]{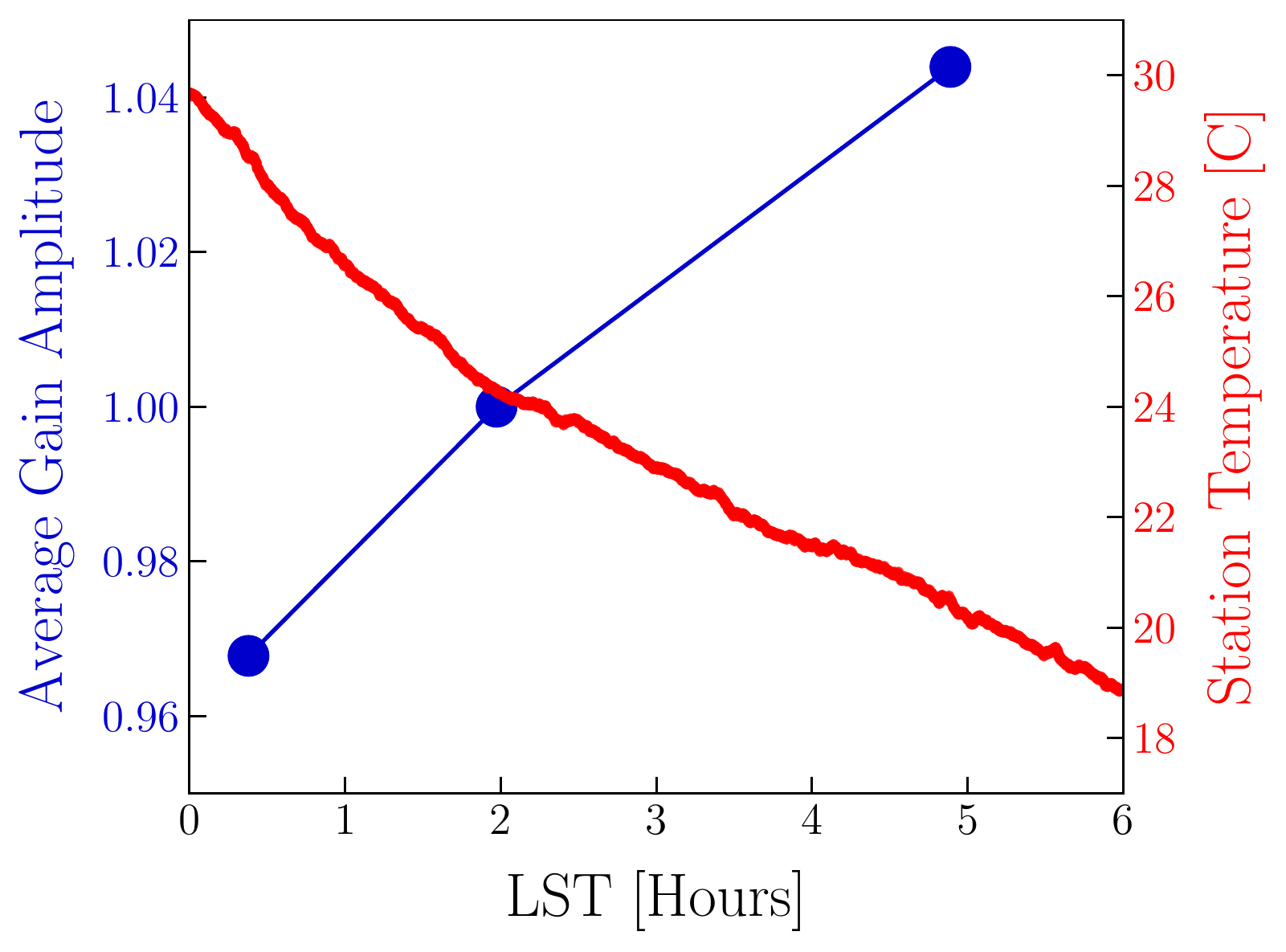}
\caption{The average gain amplitude drift (blue) throughout the 2458098 observing night, derived from three independent calibration fields and normalized to the field at 2 hours.
We also overplot the ambient temperature measured by a nearby weather station (red), showing an expected inverse correlation with the gain drift.
Using \autoref{eq:gain_temp_coeff} these data yield a gain temperature coefficient of -0.031 dB K$^{-1}$.}
\end{figure}

A steady decrease in ambient temperature after sunset can cause slow evolution in the performance of the exposed part of the signal chains, in particular the low-noise amplifier in the FEM, which is attached to the feed.
This kind of gain drift is expected to be slow but could add up over the course of an entire night of observing, especially if we choose to calibrate the data once at either the beginning or end of the night.
To test this, we calibrate a single night of data at three different fields (\autoref{fig:gsm_fields}) at different times of a single night, and compare the average gain amplitude derived from each field.
\autoref{fig:avg_gain_drift} shows this drift having normalized the gains to the 2-hour field, demonstrating a slow drift that over the course of $\sim$5 hours leads to about a 10\% drift in the gain amplitude.
Also plotted is the ambient temperature measured by a nearby weather station, which shows an expected inverse correlation with the antenna gain.
Similarly, the band-averaged gain phase drift (after taking out the cable delay) is kept to within 0.2 radians over the same time interval, but unlike the average amplitude the phase drift does not appear monotonic in time.

Using the temperature data we can derive an ambient temperature coefficient for the change in the average gain as a function of temperature difference \citep{Jacobs2013}.
We can represent a relationship for the difference in ambient temperature relative to the ratio of the derived gain response of the analog system as
\begin{align}
\label{eq:gain_temp_coeff}
10\cdot\log_{10}\left|\frac{g_{\rm new}}{g_{\rm norm}}\right|\ {\rm[dB]} = C\cdot(T_{\rm new} - T_{\rm norm})\ {\rm [K]},
\end{align}
where $T_{\rm norm}$ and $g_{\rm norm}$ are the ambient temperature and average gain amplitude at the time of gain calibration (i.e. our normalization time), $T_{\rm new}$ and $g_{\rm new}$ are the temperature and gain at any new time in LST, and $C$ is the temperature coefficient in units dB K$^{-1}$.
In \autoref{fig:avg_gain_drift}, for example, we have chosen the normalization to be at 2 hours LST.
Using the three data points from \autoref{fig:avg_gain_drift}, we derive a temperature coefficient of -0.031 dB K$^{-1}$ for the gains.
With a similar approach, \citet{Pober2012} also derive a gain temperature coefficient of -0.03 dB K$^{-1}$ for the PAPER system, which used similar front-end hardware as HERA Phase I. \citet{Jacobs2013} also used data from two different seasons to derive an auto-correlation temperature coefficient for the PAPER system of -0.06 dB K$^{-1}$, which, when divided by a factor of two in order to map it to a gain temperature coefficient, is also in agreement with these results.

%\begin{figure}
%\label{fig:gleam02_multinight_gain_comparison}
%\centering
%\includegraphics[scale=0.45]{imgs/gleam02_multinight_gain_comparison.pdf}
%\caption{Gain solution for antenna 84 from the GLEAM 02H field across four different nights, showing the inherent stability of the gains night-to-night.}
%\end{figure}

%\begin{figure*}
%\label{fig:gleam05_MFS_pI}
%\centering
%\includegraphics[scale=0.65]{imgs/gleam05_MFS_pI.pdf}
%\caption{}
%\end{figure*}
%
%\begin{figure}
%\label{fig:gleam02H05H_cal_comparison}
%\centering
%\includegraphics[scale=0.45]{imgs/gleam02H05H_cal_comparison.pdf}
%\caption{}
%\end{figure}

%\begin{figure*} 
%\label{fig:gleam02_night_imdiff}
%\centering
%\includegraphics[scale=0.60]{imgs/gleam02_night_imdiff.pdf}
%\caption{}
%\end{figure*}

%%%%%%%%%%%%%%%%%%%%%%%%%%%%%
%%%%% Combining Redundant Calibration %%%%%%
%%%%%%%%%%%%%%%%%%%%%%%%%%%%%
\section{Combining Redundant Calibration}
\label{sec:redcal}
A key part of HERA's design is to exploit its inherent redundant sampling of the $uv$ plane for precision redundant calibration \citep{Dillon2016}.
Redundant calibration asserts that all visibilities of the same baseline length and orientation (uniquely defining a ``baseline type'') measure the same visibility, which with enough redundant baselines allows for an overconstrained system of equation while keeping the true visibility a free parameter \citep{Wieringa1992, Liu2010}.
This means that redundant calibration does not need an estimate of the true model visibilities, and thus temporarily skirts some of the issues with incomplete or inaccurate sky models.
In practice this is never exactly true, and slight antenna position and primary beam uncertainties therefore generate gain errors in redundant calibration \citep{Ewall-Wice2017, Orosz2019}.
Nonetheless, we would like to explore options for combining redundant and absolute calibration to exploit their complementary advantages, either as an alternative or hybrid calibration pipeline.

For a baseline between antennas $i$ and $j$ and another between antennas $j$ and $k$, both belonging to the same baseline type of $ij$ (for example antenna pairs 23 \& 24 and 24 \& 25 from \autoref{fig:uvplot}), the redundant calibration equations are
\begin{align}
\label{eqn:redcal}
V_{ij}^{\rm data} &= g_iV_{ij}^{\rm model}g_j^\ast + n_{ij}\\
V_{jk}^{\rm data} &= g_jV_{ij}^{\rm model}g_k^\ast + n_{jk} \nonumber \\
&\ \vdots \nonumber
\end{align}
Note that the model visibility for $V_{jk}$ is now $V_{ij}^{\rm model}$.
In this case, we are left with four free parameters, $g_i$, $g_j$, $g_k$, and $V_{ij}^{\rm model}$, which we can solve for by minimizing their chi-square,
\begin{align}
\label{eqn:redcal_chisq}
\chi^2 = \sum_{i,j} \frac{|V_{ij}^{\rm data} - g_iV_{ij}^{\rm model}g_j^\ast|}{\sigma_{ij}^2},
\end{align}
where $\sigma_{ij}^2$ is the noise variance on baseline $ij$ and the sum is over all antenna pairs in the array.
Although a two-baseline array like the one in \autoref{eqn:redcal} is not redundantly calibrate-able, we can see that increasing the number of redundant baselines will turn this into an overconstrained system of equations \citep{Liu2010}.
However, redundant calibration is not the final answer for antenna-based calibration, as there exist fundamental degeneracies that redundant calibration simply cannot constrain.
One of these degeneracies is the average gain and model visibility amplitude.
Looking at \autoref{eqn:redcal_chisq}, we can see that if we multiply all antenna gains by some fraction $A$, and then divide all model visibilities by $A^2$ we leave the final $\chi^2$ unchanged.
Recall we are free to do this because, unlike in sky-based calibration, the model visibility is a free parameter.
Thus it can perfectly counteract such deviations in the gains and implies that the full system of equations is insensitive to their average amplitude.
In addition to the average gain amplitude, the other major degeneracy associated with redundant calibration is known as the ``tip-tilt'' phase gradient across the East-West and North-South coordinates of the array \citep{Zheng2014, Dillon2018}.
If each antenna is assigned a vector $\mathbf{r}_i$ originating from the center of the array to its topocentric coordinates of East \& North, we can insert a ``tip-tilt'' phase gradient into the gains as
\begin{align}
\label{eqn:TT}
g_i &\rightarrow g_i\exp(i\mathbf{\Phi}\mathbf{r}_i)\\
{\rm where}\ \mathbf{\Phi} &= \left(\Phi_E,\ \Phi_N\right). \nonumber
\end{align}
The coefficient $\mathbf{\Phi}$ is therefore a phase gradient coefficient with units of radians per meter, with separate coefficients for the East and North directions.
Such a perturbation to the gains is a degeneracy in redundant calibration because we can exactly cancel this out by applying the opposite term to the model visibilities.
For example, we can express the second term in the chi-square metric of \autoref{eqn:redcal_chisq} as
\begin{align}
g_i&V_{ij}^{\rm model}g_j^\ast \rightarrow \nonumber \\
&g_i\exp(i\mathbf{\Phi}\mathbf{r}_i)V_{ij}^{\rm model}\exp(-i\mathbf{\Phi}\mathbf{r}_{ij})g_j^\ast\exp(-i\mathbf{\Phi}\mathbf{r}_j) \nonumber \\
&= g_iV_{ij}^{\rm model}g_j^\ast\exp(i\mathbf{\Phi}\mathbf{r}_{ij})\exp(-i\mathbf{\Phi}\mathbf{r}_{ij}) \nonumber \\
&= g_iV_{ij}^{\rm model}g_j^\ast,
\end{align}
where we have made use of the fact that $\mathbf{r}_i - \mathbf{r}_j = \mathbf{r}_{ij}$, and can see that after substitutions the term is unchanged.
This amounts to a total of three parameters, the average gain amplitude and the east and north phase gradient, that need to be solved for after redundant calibration and require a sky model to pin down (per frequency, time and polarization).
Thus, the issues of inaccurate sky models are somewhat mitigated but not totally circumvented by redundant calibration \citep{Byrne2019}.
From a sky-based calibration perspective, these degeneracies can roughly be thought of as the overall flux scale of the array and its pointing center on the sky.

%\begin{figure}
%\label{fig:cal_steps}
%\centering
%\includegraphics[scale=0.10]{imgs/cal_steps.jpg}
%\caption{Three kinds of related calibration strategies including redundant + partial absolute calibration (RA Cal), sky + redundant calibration (AR Cal) and just sky calibration.}
%\end{figure}
\begin{figure}
\label{fig:sky_red_flow}
\centering
\includegraphics[scale=0.45]{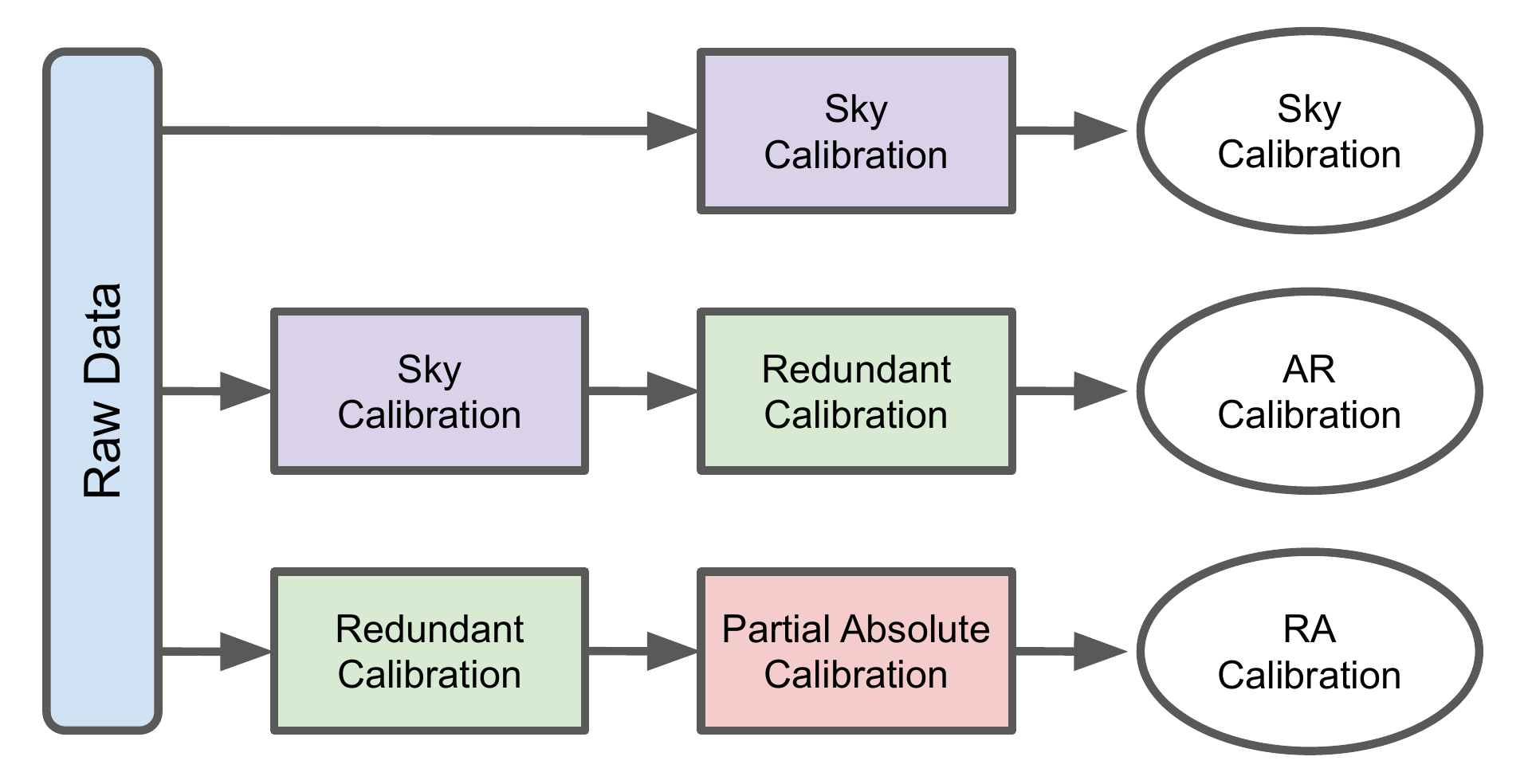}
\caption{Schematic showing the order of operations for three related calibration strategies, similar to that of \citet{Li2018}.
For AR and RA calibration, the gains from the first step are applied to the data before proceeding to the second step.
In addition, the gains derived by redundant calibration have their degenerate modes projected out before proceeding.
}
\end{figure}

\begin{figure*}
\label{fig:gleam02_abs_omni_gain_breakdown}
\centering
\includegraphics[scale=0.55]{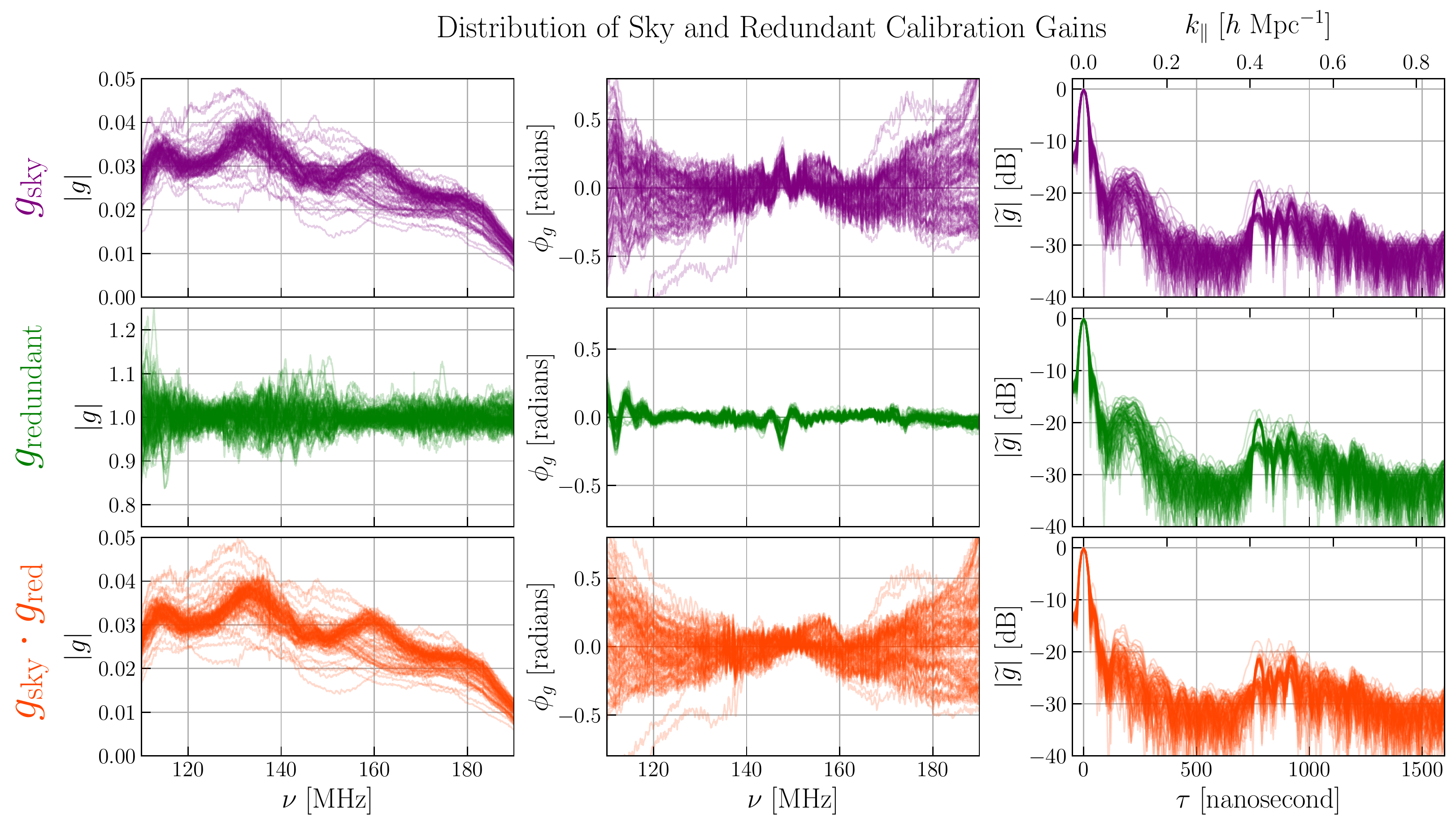}
\caption{The distribution of gain solutions from AR calibration. The top panels show just the sky calibration (similar to \autoref{fig:gleam02_raw_gains}) in amplitude, phase (after removing their cable delay) and in delay space. Middle panels show just the redundant calibration portion of the gains in AR calibration. Bottom panels show the product of the two steps, which forms the full AR calibration gain solutions. Note the notch in the phase plots that is canceled out by redundant calibration, which leads to some suppression of the 200 nanosecond feature.}
\end{figure*}

There are multiple ways to fill-in the missing degenerate parameters of redundant calibration.
One approach is to take the redundant calibration solutions and project only its degenerate components onto the degenerate modes in the sky-based calibration solutions \citep{Li2018}.
One can also take model visibilities and setup a new calibration equation that solves explicitly for the degenerate parameters (partial absolute calibration).
Finally, one can take the sky-based calibrations as a starting point by applying them to the data and then run redundant calibration.
The latter two of these, along with standard sky calibration, are shown in \autoref{fig:sky_red_flow}, outlining the order of operations of the three proposed calibration schemes: sky calibration, sky + redundant (AR) calibration and redundant + partial absolute (RA) calibration.
Both AR and RA calibration schemes are built into the \texttt{redcal} module of the \texttt{hera\_cal} software package (which we use here), including setting up and solving a system of equations that specifically picks out the degenerate parameters of redundant calibration given a set of sky model visibilities, which we discuss in more detail in \autoref{sec:partial_abs}.
For the RA approach discussed here, the model visibilities used for extracting the degenerate modes are simply the raw data calibrated with the sky-based gains.

\begin{figure*}
\label{fig:gleam02_omni_abs_sky_full_wedge}
\centering
\includegraphics[scale=0.48]{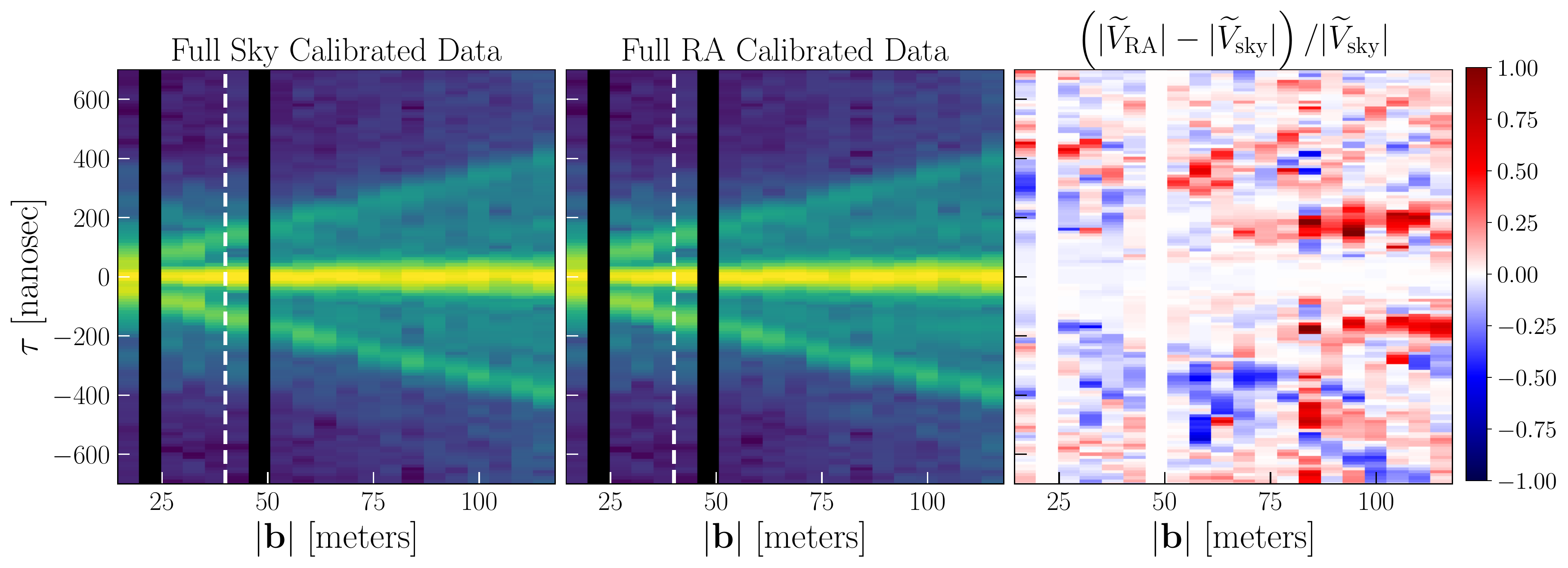}
\caption{Redundantly-averaged pseudo-Stokes I visibilities in delay space transformed across a 120 -- 180 MHz spectral window and ordered according to baseline length, having applied the full sky calibration gains (left) and the full RA calibration gains (center).
These are plotted on the same colorscale as \autoref{fig:data_model_wedge}.
Taking their fractional difference (right) shows that the RA calibration introduces less structure into the data at $|\tau|\sim200$ ns for shorter baselines (blue regions), although it also seems to introduce new structure at slightly smaller delays for long baselines (red regions).}
\end{figure*}

Because redundant calibration cannot constrain the degenerate modes inherent to its system of equations, the output gains will generally have some random combination of degenerate vectors, which will be influenced by the convergence of the calibration solver and its starting point from the raw data.
To fix this, we can \emph{project out} these degeneracies by fixing them to some a priori chosen position, which will then get filled in by absolute calibration \citep{Dillon2018, Li2018}.
The simplest thing is to re-scale the gains such that the average amplitude is 1.0 and the phase gradient is 0.0, which is done to just the redundant calibration portion of the gains in both RA and AR calibration.

We saw in \autoref{fig:gleam02_raw_gains} the presence of antenna-based structures that we expect to appear in the gains, like the dish reflection and the 20-meter and 150-meter cable reflections, but we also saw significant contamination by instrumental coupling across a wide range of delays.
To understand the kinds of structures picked up by redundant calibration we can inspect the gains in a similar manner.
\autoref{fig:gleam02_abs_omni_gain_breakdown} shows the distribution of the gains at each step in the AR calibration scheme in amplitude (left) and in phase (right) having removed the cable delay for each antenna.
It also shows the gains Fourier transformed across frequency in delay space (right) and peak-normalized.
The top panels of \autoref{fig:gleam02_abs_omni_gain_breakdown} show just sky calibration (the same as \autoref{fig:gleam02_raw_gains}).
The middle panel shows just the redundant calibration component of the gain, where in deriving them we first apply the sky calibration gains to the data, and the bottom panels shows the final product of the two gains.
Note that the redundant calibration gains derived here use the same baselines as the sky calibration of $|\mathbf{b}| > 40$ meters.
For the redundant calibration gains, we can see that its average amplitude is one as expected, and has similar kinds of spectral structure as the sky calibration gains.
Looking at their product, or the AR calibration gains, we can see some of the benefits of redundant calibration.
Compared to the sky calibration delay response (purple), the AR delay response (orange) has a slightly suppressed bump at $\sim$200 ns, which can also be seen as the negation of the coherent ripple in the center phase plots.
We observe this ripple in the sky-based gain phases (top-center), which seems to be corrected-for by redundant calibration (middle-center) such that their product (bottom-center) demonstrates less of a ripple.
One possible explanation is that this ripple is caused by an imperfect sky model that creates spectral errors in the sky-based gain that is then corrected by redundant calibration.
However, we still see significant power at $\tau\gtrsim200$ ns, which could originate from non-redundancies between nominally redundant baselines specifically at the horizon, where diffuse emission generates the pitchfork effect in the data but also where the per-antenna primary beams are likely the least redundant with each other.
Similar to how unmodeled diffuse emission created gain errors in sky calibration, these kinds of non-redundancies will create errors in redundant calibration and will appear at similar delays \citep{Orosz2019}.
Previous work showed that non-redundancy seems to be worse for short baselines \citep{Carilli2018}, but quantifying this in more detail is still in progress \citep{Dillon2019}.
The AR calibration gains also show significant power at $\tau\gtrsim800$ ns, which shows that redundant calibration is not immune to picking up cross coupling instrumental systematics.
The RA gain solutions show nearly the same structure as solutions derived from AR calibration down to below 1\% in fractional difference, so we do not plot them here for brevity.

To further show the effects of redundant calibration on the data, we take the full gains (this time from the RA calibration scheme) and apply them to the data.
We then redundantly average and Fourier transform them similar to \autoref{fig:data_model_wedge}.
\autoref{fig:gleam02_omni_abs_sky_full_wedge} shows this process having applied the full sky calibration and full RA calibration, and also shows the fractional difference between the two (right).
Areas where the RA calibration is introducing new structures show up as red, and areas where sky calibration is introducing structure where RA is not show up at blue.
We see that for shorter baselines and delays near $|\tau|\sim200$ ns, that RA calibration is inserting less structure into the data compared to sky calibration, which agrees with our observation earlier that spurious gain structure at those delay scales seem suppressed.
However, we see that at slightly smaller delays and larger baseline lengths, RA calibration is also inserting additional power compared to sky calibration, which is likely a result of its own gain errors.
Additionally, at small delays ($|\tau| < 100$ ns) the two are in good agreement with each other. 

%\begin{figure*}
%\label{fig:gleam02_abs_omni_mfs}
%\centering
%\includegraphics[scale=0.60]{imgs/gleam02_abs_omni_mfs.pdf}
%\caption{}
%\end{figure*}

The take-away from this section is: 1) all three calibration schemes yield gains that are similar at low delays; 2) hybrid redundant calibration seems to correct for some of the errors in the sky-based calibration but still introduces its own set of errors; 3) both sky and redundant calibration suffer from gain errors that are induced by baseline-dependent instrumental systematics.
Moving forward, future analyses will benefit from attempting to model diffuse emission and removing instrumental cross coupling systematics before calibration in order to calibrate intermediate delay scales and exploit the full power of a combined redundant and absolute calibration approach.

\citet{Li2018} performed a similar comparison with the MWA, using the Fast Holographic Deconvolution package \citep[FHD;][]{Sullivan2012} for sky-based calibration and the \texttt{omnical} package \citep{Zheng2014} for redundant calibration.
Similar to this work, they find marginal improvements with a combined sky + redundant calibration approach.

%%%%%%%%%%%%%%%%%%%%%%%%%%%%%
%%%%%% Power Spectrum Performance %%%%%%%
%%%%%%%%%%%%%%%%%%%%%%%%%%%%%
\section{Power Spectrum Performance}
\label{sec:pspec}
We use the visibility-based, delay spectrum estimator of the \tocm power spectrum to further assess the quality of the calibration and the overall stability of the array.
The delay transform is simply the Fourier transform of the visibilities across frequency into the delay domain
\begin{align}
\label{eq:delay_transform}
\widetilde{V}(\mathbf{u}, \tau) = \int d\nu\ e^{2\pi i\nu\tau} V(\mathbf{u}, \nu),
\end{align}
where $\mathbf{u} = \mathbf{b} / \lambda$ is the $uv$ vector of the baseline and $\lambda$ is the observing wavelength \citep{Parsons2012b, Liu2014a, Parsons2014}.
The Fourier dual of frequency, $\tau$, is not a direct mapping of the line-of-sight spatial wavevector $k_\parallel$ but under certain assumptions it is a fairly good approximation.
This is known as the ``delay approximation'' and has been shown to be fairly accurate for short baselines \citep{Parsons2012b}.
The delay spectrum estimate of the \tocm power spectrum is the delay transformed-visibilities squared, multiplied by the appropriate scaling factors,
\begin{align}
\label{eq:delay_spectrum}
\widehat{P}_{21}(\kperp, \kpara) \approx |\widetilde{V}(\mathbf{u}, \tau)|^2\frac{X^2Y}{\Omega_{pp}B_{p}}\left(\frac{c^2}{2k_{B}\bar{\nu}^2}\right)^2,
\end{align}
where $X$ and $Y$ convert angles on the sky and delay modes to cosmological length scales, $\Omega_{pp}$ is the sky-integral of the squared primary beam, $\bar{\nu}$ is the average frequency in the delay transform window and $B_{p}$ is the delay transform bandwidth, as defined in Appendix B of \citet{Parsons2014}.
The relationships between the Fourier domains inherent to the telescope, $\mathbf{u}$ and $\tau$, and the cosmological Fourier domains are
\begin{align}
\label{eq:cosmo_scalings}
\kpara &= \frac{2\pi}{X}\tau \nonumber \\
\kperp &= \frac{2\pi}{Y}\frac{b}{\lambda},
\end{align}
where $X = c(1+z)^2\nu_{21}^{-1}H(z)^{-1}$, $Y = D(z)$, $\nu_{21}=1.420$ GHz, $H(z)$ is the Hubble parameter, $D(z)$ is the transverse comoving distance, $b$ is the baseline length and $\lambda$ is the observing wavelength \citep{Parsons2012a, Liu2014a}.
For this analysis, we adopt a $\Lambda$CDM cosmology with parameters derived from the \emph{Planck} 2015 analysis \citep{Planck2016}, namely $\Omega_\Lambda=0.6844$, $\Omega_b=0.04911$, $\Omega_c=0.26442$ and $H_0=67.27$ km/s/Mpc.

Due to the chromaticity of an interferometer, foreground emission that is inherently spectrally smooth (such as galactic synchrotron) will have increased spectral structure in the measured visibilities.
The delay at which the instrument imparts this spectral structure is dependent on the geometric delay of the source signal between the two antennas that make up a baseline, given as
\begin{align}
\label{eq:wedge}
\tau = \frac{|\mathbf{b}|\sin(\theta)}{c},
\end{align}
where $\theta$ is the zenith angle of the incident foreground emission and $\mathbf{b}$ is the baseline separation vector.
We can see that spectrally-smooth foregrounds incident from zenith will appear at lower delays and therefore have less induced chromaticity, while foregrounds incident from large zenith angles will have more induced chromaticity.
The maximum delay a smooth spectrum foreground can appear at is called the horizon limit, in which case $\tau_{\rm horizon} = \tau(\theta=90^\circ)$.
If we could perfectly image the interferometric data we could also reconstruct the smooth spectrum foregrounds.
However, this is in practice never the case, as effects like missing $uv$ samples and imaging via gridded Fourier transforms create low-level chromatic sidelobes that corrupt the images with spectrally-dependent residual foregrounds.
Visibility-based power spectrum estimators that do not even attempt to image the data are stuck with the most severe amounts of instrument-induced chromaticity, generally out to the baseline horizon delay.
The horizon limit is a function of baseline length (\autoref{eq:wedge}), and as such it forms a wedge-like shape in the data's Fourier domain and has come to be known as the foreground wedge \citep{Datta2010, Morales2012, Parsons2012b, Thyagarajan2013, Liu2014a, Morales2019}.
Because HERA has a fairly compact primary beam we expect most foreground power to lie within $\tau \le |\mathbf{b}|\sin(\theta=5^\circ)/c$; however, the vast amounts of diffuse emission near the horizon means that we still expect to see some amount of foreground power out to the horizon limit, even though it is significantly attenuated by the primary beam \citep{Thyagarajan2016}.

%\begin{figure}
%\label{fig:flag_mask}
%\centering
%\includegraphics[scale=0.60]{imgs/flag_mask.pdf}
%\caption{Visibility flagging mask used to excise data contaminated with RFI and other spurious aftifacts. The large excised bandwidth at 137 MHz is due to the bright ORBCOMM communication satellites.}
%\end{figure}

\begin{figure*}
\label{fig:wedge_2458098_PX_abs_smooth_autobl_linpol}
\centering
\includegraphics[scale=0.6]{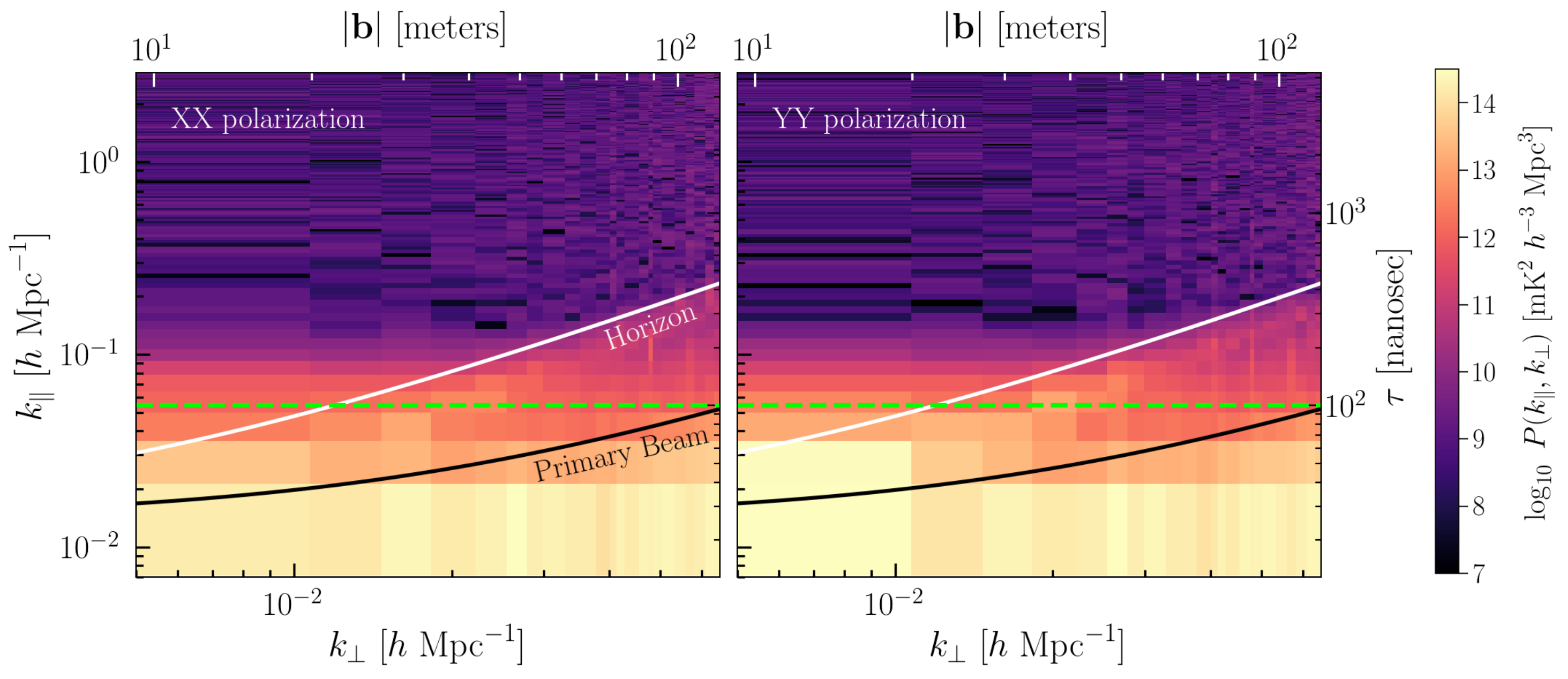}
\caption{Wide-band, two-dimensional power spectra of each linear dipole polarization XX (left) and YY (right) having applied the smooth sky-based calibration, and after systematic removal and an incoherent average (i.e. after squaring the visibilities) from 0 to 2 hours LST.
Power spectra are formed between 139 -- 178 MHz having applied a Blackman window to limit spectral leakage in the discrete Fourier transform.
The black line marks the FWHM of the primary beam ($\pm5^\circ$ from zenith) and the white line marks the baseline horizon.
Both lines have an additive buffer of $k_\parallel=0.014\ h\ {\rm Mpc^{-1}}$ to account for the width of the Blackman kernel in Fourier space.
The dashed green line marks the maximum delay scale of the smoothed gain solutions.
Most of the foreground power is confined within the horizon limit of the array, however there is evidence for some supra-horizon leakage at short baselines.}
\end{figure*}

The issue of whether foregrounds actually appear tightly confined within the foreground wedge is an open question: 21\,cm foreground studies seem to indicate that \emph{supra-horizon} foreground power tends to extend only slightly beyond the horizon \citep{Pober2013b, Bernardi2013, Gehlot2018, Lanman2019c}, but whether this is truly the case down to EoR sensitivities is not known.
There are a number of effects that can contribute to measured supra-horizon foreground emission, including intrinsic foreground spectral structure, unflagged RFI, primary beam chromaticity, and also gain calibration errors.
As discussed in \autoref{sec:gains}, the intrinsic gain kernel of the instrument may have a non-negligible extent to large delay modes, which if left uncalibrated will push foreground power out to higher delays.
Similarly, gain errors will introduce structure at these scales and have the same effect.
Smoothing the gains eliminates the latter concern but still leaves the possibility of the former effect.
To asses the degree of foreground containment we can form wide-band, visibility-based power spectra as a diagnostic.

This is complemented by an understanding of how thermal noise appears in the power spectra.
Given our knowledge of the noise properties of our antennas, we can compute a theoretical estimate of the noise power spectrum, $P_{\rm N}$, which is equivalent to the root-mean square (RMS) of the power spectrum if the only component in the data were noise.
This is one measure of the uncertainty on the power spectra due to noise, but also represents the theoretical amplitude of the power spectra in the limit that they are noise dominated (as opposed to signal or systematic dominated).
This is given in \citet{Cheng2018} as
\begin{align}
\label{eqn:PN}
P_{\rm N} = \frac{X^2Y\Omega_{\rm eff}T_{\rm sys}^2}{t_{\rm int}N_{\rm coherent}\sqrt{2N_{\rm incoherent}}},
\end{align}
where the $X$ and $Y$ scalars are the same as before, $T_{\rm sys}$ is the system temperature in milli-Kelvin, $t_{\rm int}$ is the correlator integration time in seconds, $N_{\rm coherent}$ is the number of sample averages done at the visibility level (i.e. before visibility squaring), and $N_{\rm incoherent}$ is the number of sample averages done at the power spectrum level (i.e. after visibility squaring).
$\Omega_{\rm eff}$ is the effective beam area given by $\Omega_{\rm eff} = \Omega_{p}^2 / \Omega_{pp}$, where $\Omega_{p}$ is the integral of the beam across the sky in steradians, and $\Omega_{pp}$ is the integral of the squared-beam across the sky in steradians \citep{Pober2013b, Parsons2014}.
Using similar data products, \citet{Kern2019b} showed that the HERA Phase I system achieves an antenna-averaged $T_{\rm sys}\sim250$ K at 160 MHz, which we adopt in this work.

\begin{figure*}
\label{fig:gleam02_PX_abs_red_pspec}
\centering
\includegraphics[scale=0.50]{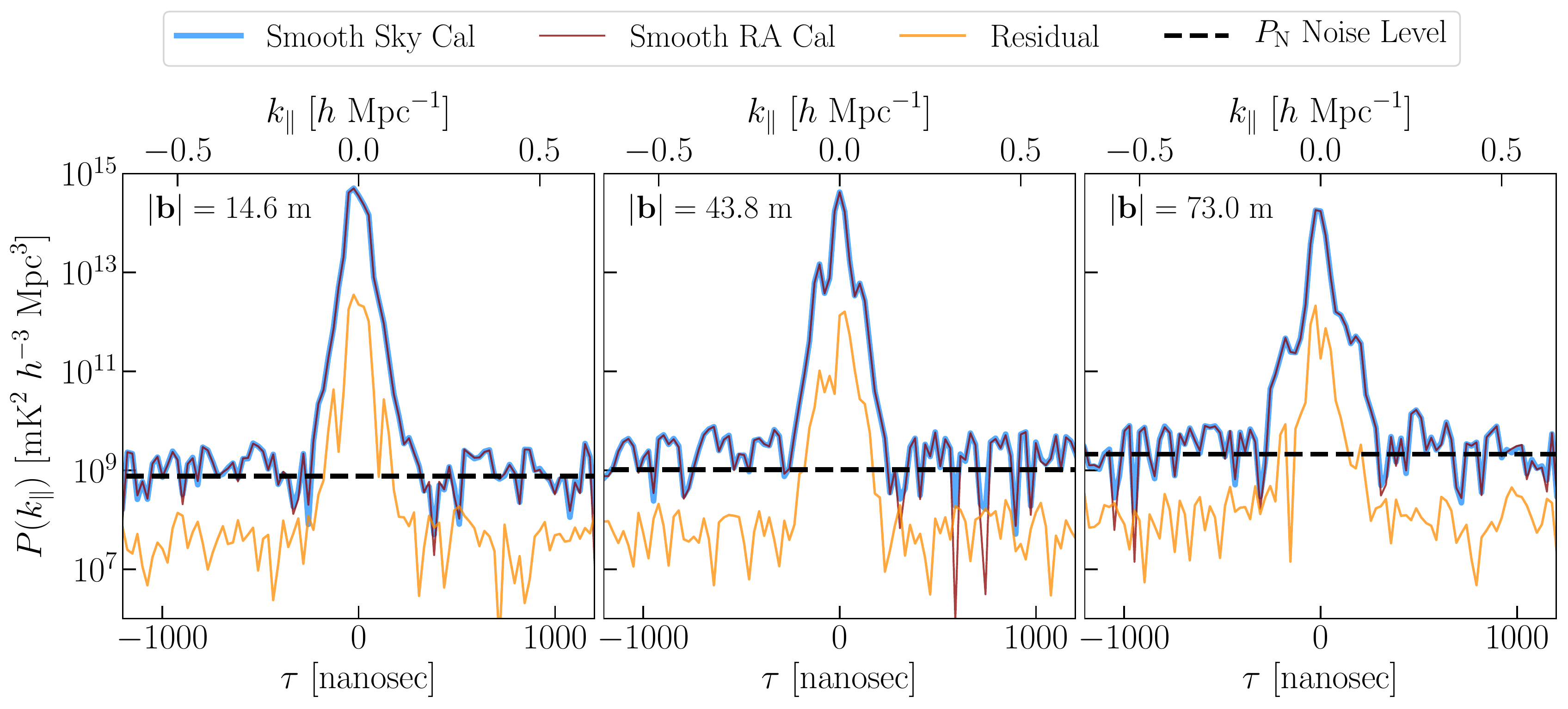}
\caption{Delay spectra of three redundantly averaged East-West baseline types for the instrumental YY polarization, showing the data calibrated with the smooth sky calibration (blue), the smooth RA calibration (red) and their residual (red), along with the thermal noise floor (dashed-black) assuming a $\rm T_{\rm sys} = 250$ K. The two calibration yield nearly the same averaged power spectra across all delays, which show consistency with the theoretical noise floor outside $k_\parallel \gtrsim 0.2\ h\ {\rm Mpc}^{-1}$.}
\end{figure*}

The raw data are flagged for radio frequency interference (RFI) and are thus nulled at the flagged channels.
This leads to a highly discontinuous windowing function that when taking a Fourier transform will spread foreground power and contaminate the EoR window.
To prevent this, we employ the same 1D delay domain deconvolution as the gain smoothing filter (\autoref{sec:gains}) on each visibility, filling in model CLEAN components out to 2000 nanoseconds.
HERA Phase I data are contaminated by cable reflection and cross coupling instrumental systematics \citep{Kern2019b}.
Because we are concerned with foreground leakage due to calibration in this work, we remove these systematics before forming power spectra for visual clarity.
Specifically, we apply a time-domain filter to suppress cross coupling in all visibilities with a projected East-West length greater than 14 meters (throwing out all other visibilities).
This time-domain filter is performed in the delay domain, and isolates a rectangle spanning $|\tau| > 0.8 \tau_{\rm horizon}$ and fringe-rates given by the 99\% EoR power bounds in \citet{Kern2019a} for each baseline independently.
We also calibrate out a single cable reflection term for each of the 20-meter and 150-meter cables in the analog system per dipole polarization also using the methods in \citet{Kern2019a}.

We form power spectra across a spectral window from 139 -- 178 MHz and apply a Blackman window prior to taking the Fourier transform to limit spectral leakage in the discrete Fourier transform.
Baselines are only cross-multiplied with themselves, and are not cross-multiplied with other baselines in a redundant group.
Normally this would produce a noise-bias in the power spectra, so instead we cross-multiply baselines with themselves at adjacent time integrations, having first rephased them to the same pointing center \citep{Pober2013b}.
We do this for all baselines for each time integration pair in the range of 0 -- 2 hours LST.
After squaring the visibilities, we incoherently average the power spectra across LST and then average all power spectra of the same baseline length (regardless of orientation), which is equivalent to cylindrically gridding $\mathbf{k}$ space into $k_\perp$ and $k_\parallel$ annuli.

\autoref{fig:wedge_2458098_PX_abs_smooth_autobl_linpol} shows the 2D power spectra in instrumental XX and YY visibility polarizations with the smooth sky calibration gains applied to the data (smoothed out to $\tau=100$ ns).
We also show the primary beam FWHM limit (black) and the full horizon limit (white) in both instrumental XX and YY visibility polarization.
Both lines have an additive buffer of $k_\parallel = 0.014\ h\ {\rm Mpc}^{-1}$ to account for the width of the Blackman kernel in Fourier space.
The dashed green line shows the maximum delay scale of the applied gains after smoothing.
We find that most of the foreground power is contained within the horizon limit, with some amounts of supra-horizon leakage for short baselines.
The strong pitchfork feature of the foreground emission tracing the horizon line is not as prominent in this plot as it was in \autoref{fig:data_model_wedge}, which is due to the fact that it was partially removed in the cross coupling filter applied to the data.
\citet{Kern2019b} showed that the edges of the pitchfork are slowly time variable and thus can be separated from the cosmological 21\,cm signal and filtered out with a high-pass time filter.
\autoref{fig:gleam02_fullcal_dly_frate} also demonstrates this, showing that the two lobes at $\pm\tau=100$ ns are also centered at $f=0$ mHz, meaning they primarily contain slowly time-variable terms.
This means that the time filter designed to eliminate cross coupling also helps to reduce some of the strongest foreground emission straddling the boundary of the foreground wedge and EoR window.

To first order, \autoref{fig:wedge_2458098_PX_abs_smooth_autobl_linpol} tells us that our single field, smoothed sky-based calibration with restricted degrees of freedom has done a fairly good job calibrating the data and has largely kept foreground power contained within the foreground wedge.
For short baselines, however, we can begin to see evidence for some amount of supra-horizon emission that could be due to uncalibrated gain terms or imperfectly removed cross coupling, the latter of which is harder to remove for shorter baselines.
This supra-horizon emission is located beyond the smoothing scale of the gains and appears in amplitude slightly larger than predictions of the high-order dish reflections \citep{Patra2018}, but is contained within $k_\parallel \lesssim 0.2\ h\ {\rm Mpc}^{-1}$ down to nearly $\sim10^6$ in dynamic range against the foreground peak.
Note that, possibly coincidentally, this supra-horizon excess seems more prevalent for baselines shorter than our initial baseline cut of 40 meters.
Deeper integrations will help to discriminate whether the observed supra-horizon emission extends further out in $k_\parallel$ space at lower noise levels.

\autoref{fig:gleam02_PX_abs_red_pspec} shows the same power spectra but focuses on three unique baseline types: purely East-West baselines of 14.6 m, 43.8 m, and 73 m in length.
In addition to showing power spectra of the data with the smooth sky calibration (blue), we also show the smooth RA calibration (also smoothed out to $\tau=100$ ns) and the residual between the two.
This demonstrates that the calibration strategies, post-smoothing, have nearly the same impact on the averaged power spectrum.
We also show the theoretical noise-floor of the data (dashed-black), which more clearly demonstrates the agreement of the data with the noise floor outside $k_\parallel \gtrsim 0.2\ h\ {\rm Mpc}^{-1}$.
Note that the noise floor for longer baseline types is higher because there are fewer physical baselines, meaning less averaging is done in the (coherent) redundant average.

%%%%%%%%%%%%%%%%%%%%%%%%%%%%%
%%%%%%%%%%%% Summary %%%%%%%%%%%%
%%%%%%%%%%%%%%%%%%%%%%%%%%%%%

\section{Summary}
\label{sec:summary}
In this work we discuss sky-based and hybrid-redundant calibration strategies for Phase I of the Hydrogen Epoch of Reionization Array.
We present a CASA-based calibration pipeline for constructing sky models, applying primary beam corrections and performing direction-independent, antenna-based complex gain calibration.
We use this to characterize the time and frequency stability of the gain solutions, finding that slow and fast nightly time drifts in the gain's overall flux scale are order 8\% and 0.1\% effects, respectively.
We also show that unmodeled diffuse foregrounds, mutual coupling and other cross coupling systematics in the HERA Phase I system are picked up in the process of both sky and redundant calibration, and limit the accuracy of our gain calibration at intermediate delay scales $\tau \gtrsim 100$ nanoseconds.
By low-pass filtering the gains with a Fourier filter, we can restrict the degrees of freedom of the gains and we show that this mitigates the effect of these gain errors.
Additionally, while we do not perform any kind of polarization calibration, we find that polarization leakage from Stokes I to Q, U \& V is on the order of a few percent.

We also present a hybrid approach for combining redundant and absolute calibration techniques and, similar to \citet{Li2018}, find that the hybrid technique marginally improves the gain solutions over just sky-based calibration, although we did not perform any self-calibration iterations which would likely have improved the fidelity of the sky-based gains.
This was omitted because of the difficulty of implementing this to high dynamic range given the mediocre Phase I angular resolution, but will be explored in future work.
Additionally, we show that the hybrid calibration scheme is also limited by gain errors at similar $\tau\gtrsim100$ ns scales as the sky calibration, which we suggest can be further mitigated by enacting a larger minimum baseline cut in the calibration as well as attempts to include the diffuse emission component of the sky into the calibration model.

Finally, we form two-dimensional power spectra across 139 -- 178 MHz from 0 -- 2 hours LST, and show that most of the foreground power measured by HERA is contained within the horizon limit of the array, but we do observe non-negligible supra-horizon power for short baselines that were not included in the calibration.
This emission is confined within $k_\parallel \lesssim 0.2\ h\ {\rm Mpc}^{-1}$ down to the noise floor of the data, which achieves a dynamic range of nearly $10^6$ against the peak foreground power.
This could be due to uncalibrated gain terms at these scales or also from residual instrumental systematics.
Deeper integrations will help make this clearer and will help us understand how far in $k_\parallel$ the supra-horizon emission extends.
In repeating the analysis for both the smooth sky calibration and smooth hybrid calibration, we find that they have nearly the same impact on the power spectra.
Future observing seasons with the full HERA array will make high dynamic range imaging and direction-dependent calibration easier to implement and may be a way to mitigate some of the errors observed in the gain solutions.
Overall, our work shows that HERA Phase I can be relatively well-calibrated for a foreground avoidance power spectrum estimator with only a few degrees of freedom across the time and frequency axes.

\

This material is based upon work supported by the National Science Foundation under Grant Nos. 1636646 and 1836019 and institutional support from the HERA collaboration partners.
This research is funded in part by the Gordon and Betty Moore Foundation.
HERA is hosted by the South African Radio Astronomy Observatory, which is a facility of the National Research Foundation, an agency of the Department of Science and Innovation.
J. S. D. gratefully acknowledges the support of the NSF AAPF award \#1701536.
A. Liu acknowledges support  from a Natural Sciences and Engineering Research Council of Canada  (NSERC) Discovery Grant and a Discovery Launch Supplement, as well as the Canadian Institute for Advanced Research (CIFAR) Azrieli Global Scholars program.
Parts of this research were supported by the Australian Research Council Centre of Excellence for All Sky Astrophysics in 3 Dimensions (ASTRO 3D), through project number CE170100013.
G. B. acknowledges funding from the INAF PRIN-SKA 2017 project 1.05.01.88.04 (FORECaST), support from the Ministero degli Affari Esteri della Cooperazione Internazionale - Direzione Generale per la Promozione del Sistema Paese Progetto di Grande Rilevanza ZA18GR02 and the National Research Foundation of South Africa (Grant Number 113121) as part of the ISARP RADIOSKY2020 Joint Research Scheme, from the Royal Society and the Newton Fund under grant NA150184 and from the National Research Foundation of South Africa (grant No. 103424).

%%%%%%%%%%%%%%%%%%%%%%%%%%%%%
%%%%%%%%%%%% Appendix A %%%%%%%%%%%
%%%%%%%%%%%%%%%%%%%%%%%%%%%%%

\appendix{}
\section{Partial Absolute Calibration}
\label{sec:partial_abs}
Partial absolute calibration is the process of taking a set of sky model visibilities and setting up a system of equations that solves for just the degenerate components of redundant calibration.
The number of degenerate modes in redundant calibration depends on the kind of redundant calibration being employed \citep{Dillon2018}.
Here we discuss the degeneracies associated with the ``2-pol'' scheme, which calibrates the X and Y dipoles separately and ignores cross-feed polarization terms.
As shown in \autoref{sec:redcal}, there are three main degeneracies in redundant calibration \emph{for each dipole polarization}: the average gain amplitude (or the absolute flux scale of the instrument) and a ``tip-tilt'' phase gradient as a function of distance from the center of the array for both the East and North spatial axes (or the overall pointing center of the instrument).
Each of these parameters has an arbitrary frequency dependence, meaning that various kinds of spectral structure can occupy these degenerate modes.
We can express these parameters in the $i$th antenna gain of the X dipole as
\begin{align}
\label{eqn:partial_gains}
g_{i,X}(\nu) = \exp(\eta_{{\rm abs}, X}(\nu) + i2\pi\nu(T_{E, X}r_{i,E} + T_{N, X}r_{i,N}) +  i(\Phi_{E,X}(\nu) r_{i,E} + \Phi_{N,X}(\nu) r_{i,N})),
\end{align}
where $r_{i,E}$ is the East distance of antenna $i$ from the center of the array in meters and we have explicitly included the frequency dependence of the gain and its parameters.
Note that we have redefined the phase component into the sum of two terms, a spatial delay gradient $\mathbf{T}_{X} = (T_{E,X}, T_{N,X})$, and a spatial phase gradient $\mathbf{\Phi}_{X} = (\Phi_{E,X}, \Phi_{N,X})$.
Note that the delay gradient parameter has units of nanoseconds / meter and is itself frequency-independent, but has the effect of creating a phase slope in the gain across frequency.
The delay gradient manifests as a delay plane across the array that sets the phase center.
It forms a subspace of the original phase gradient space so we simply pull it out and redefine the phase gradient term $\Phi$ to be a deviation about the delay plane.
This is important because when we go to solve the calibration equation we want the phase measurements to be near zero or at least considerably less than $2\pi$ to mitigate phase wrapping  \citep{Liu2010}.
Phase wrapping creates local minima that confuse the calibration phase solver, which can be alleviated through pre-conditioning of the system of equations by first solving for and eliminating the delay gradient term.

Using a logarithm to linearize the calibration equation, the average antenna amplitude for the X dipole is found by solving the following system of equations
%\begin{align}
%\label{eqn:partial_abs_cal}
%\ln\left|\frac{V_{ij, XX}^{\rm data}}{V_{ij, XX}^{\rm model}}\right| = 2\eta_{{\rm abs}, X};\ \ln\left|\frac{V_{jk, XX}^{\rm data}}{V_{jk, XX}^{\rm model}}\right| = 2\eta_{{\rm abs}, X};\ \ldots\ {\rm all\ antenna\ pairs}
%\end{align}
%We can cast this into a standard linear model form as
\begin{align}
\label{eqn:avg_amp_lineqn}
\mathbf{y} = \left(\begin{array}{c}
\ln\left|\frac{V_{ij, XX}^{\rm data}}{V_{ij, XX}^{\rm model}}\right| \\[.3cm]
\ln\left|\frac{V_{jk, XX}^{\rm data}}{V_{jk, XX}^{\rm model}}\right| \\[.3cm]
\vdots
\end{array}\right) = \mathbf{A} \hat{\mathbf{x}} = 
\left(\begin{array}{c}2\\2\\ \vdots\end{array}\right) \left(\begin{array}{c}\hat{\eta}_{{\rm abs},X}\end{array}\right),
\end{align}
where we have specified now that the visibilities are from the $XX$ instrumental polarization.
We use the linear and non-linear equation solving package \texttt{linsolve} to solve these systems of equations.

The delay gradient parameter can be isolated by taking the phase of the data-model ratio:
\begin{align}
{\rm angle}\left(\frac{V_{ij, XX}^{\rm data}}{V_{ij, XX}^{\rm model}}\right)(\nu) = 2\pi\nu\mathbf{T}_{X}\mathbf{r}_{ij},
\end{align}
where the ${\rm angle}(\times)$ operator is $\tan^{-1}({\rm Im}(\times) / {\rm Re}(\times))$.
However, we can see that the delay gradient parameter is not inherently a function of frequency, so instead of solving this equation at each frequency we should re-cast it in a form that is frequency independent and then solve that equation.
This can be expressed as
\begin{align}
{\rm delay}\left(\frac{V_{ij, XX}^{\rm data}}{V_{ij, XX}^{\rm model}}\right) = \mathbf{T}_{X}\mathbf{r}_{ij},
\end{align}
where the ${\rm delay(\times)}$ operator takes the Fourier transform of its argument and identifies the delay of its peak in amplitude via quadratic interpolation of the three strongest Fourier modes.
The system of equations for the delay gradient of the X dipole is then
\begin{align}
\label{eqn:delay_gradient}
\mathbf{y} = \left(\begin{array}{c}
{\rm delay}\left(\frac{V_{ij, XX}^{\rm data}}{V_{ij, XX}^{\rm model}}\right) \\[.3cm]
{\rm delay}\left(\frac{V_{jk, XX}^{\rm data}}{V_{jk, XX}^{\rm model}}\right) \\[.3cm]
\vdots
\end{array}\right) = \mathbf{A} \hat{\mathbf{x}} = 
\left(\begin{array}{cc}r_{ij,E} & r_{ij,N}\\r_{jk,E} & r_{jk,N}\\ \vdots\end{array}\right) \left(\begin{array}{c}\hat{T}_{E,X} \\ \hat{T}_{N,X}\end{array}\right).
\end{align}
The estimated delay gradient gain is then expressed as $\hat{g}_{i, X}(\nu) = \exp(i2\pi\nu\hat{\mathbf{T}}_{X}\mathbf{r}_i)$.

After dividing the data visibilities by the estimated delay gradient gains, we can solve for the leftover phase gradient parameter for the X dipole with the system of equations
\begin{align}
\label{eqn:phase_gradient}
\mathbf{y} = \left(\begin{array}{c}
{\rm angle}\left(\frac{\widetilde{V}_{ij, XX}^{\rm data}}{V_{ij, XX}^{\rm model}}\right) \\[.3cm]
{\rm angle}\left(\frac{\widetilde{V}_{jk, XX}^{\rm data}}{V_{jk, XX}^{\rm model}}\right) \\[.3cm]
\vdots
\end{array}\right) = \mathbf{A} \hat{\mathbf{x}} = 
\left(\begin{array}{cc}r_{ij,E} & r_{ij,N}\\r_{jk,E} & r_{jk,N}\\ \vdots\end{array}\right) \left(\begin{array}{c}\hat{\Phi}_{E,X} \\ \hat{\Phi}_{N,X}\end{array}\right),
\end{align}
where $\widetilde{V}^{\rm data}$ denotes the fact that we have first divided the data visibilities with the delay gradient gain (or equivalently multiplied the model visibilities by the delay gradient gain).
The average amplitude parameter, being orthogonal to the phase parameters, does not necessarily need to precede these steps.
The full partial absolute calibration gain is then simply \autoref{eqn:partial_gains} filled with our estimates of the degenerate parameters.

\begin{figure*}
\label{fig:phase_slope}
\centering
\includegraphics[scale=0.48]{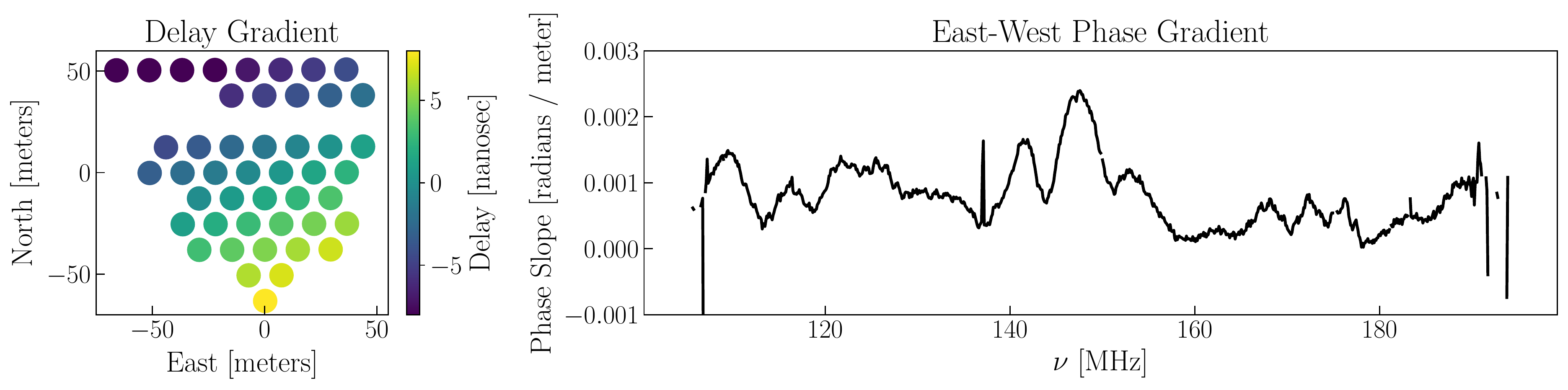}
\caption{The delay gradient (left) and East-West phase gradient (right) derived by partial absolute calibration for the X dipole polarization using the GLEAM-02H field flux density model.
We observe a significant amount of spectral structure in the phase gradient parameter, meaning it cannot be overlooked in partial absolute calibration.}
\end{figure*}

One interesting feature about delay and phase gradient calibration is that they do not require a reference antenna.
Because phase is a periodic coordinate system, sky-based phase calibration requires that we select a reference antenna whose phase is identically zero, which is a way to constrain the overall phase parameter which does not have a physical meaning.
If the array coordinates are defined in East-North-Up coordinates, then for delay and phase gradient calibration we can change the overall phase of the gain solutions by moving the estimated delay plane or phase plane up or down along the Up axis (i.e. the z-axis if the array is defined in X-Y-Z coordinates).
Moving the delay and phase planes up or down does not change the relative delay and phase between antennas, which is really what we care about.
We can pin this free parameter by selecting the Up-axis intercept of the plane, which is equivalent to setting the origin of the $\mathbf{r}_i$ vector coordinates that, up to now, we have defined as the center of the array.
We can see now that setting it at the center of the array is not strictly required, but it does make computations easier if done so.
Therefore, the act of setting the origin of the antenna position coordinate system plays the same role in delay and phase gradient calibration as the reference antenna does in standard sky-based phase calibration.

In \autoref{fig:phase_slope} we show the derived delay gradient (left) and phase gradient (right) parameters across the array for the X dipole polarization.
The phase gradient terms shows significant amounts of spectral structure, highlighting its ability to pickup on non-trivial spectral terms in the data.
For a large, 350+ element array these steps may take too long to calibrate the data in real time using all $\sim N^2$ baselines.
However, for partial absolute calibration we may get away with only using some of the baselines instead of all of them.
The degenerate parameters of redundant calibration outlined above are not actually specific to any individual antenna in the array: they are only properties of the array itself.
We could, for example, use only a fraction of the longest baselines in the array for partial absolute calibration, which gives us a lever-arm advantage for estimating the delay and phase gradient terms.
Concern about this approach are 1) increased noise in the gains due to less data points in our $\mathbf{y}$ vector and 2) if the baselines selected are drawn from a unique population of antennas relative to all other antennas in the array, in which case the average amplitude and phase gradients estimated with the longest baselines (which will preferably come from antennas near the edge of the array) will be mis-estimates for the other antennas not represented in the system of equations.
One could devise strategies for mitigating these kinds of concerns by, say, ensuring that while only a fraction of the baselines are used in calibration, every antenna is at least somewhat represented in the system of equations.

\section{Software}
\label{appendix:software}
The analysis presented in this work relies heavily on the Python programming language (\url{https://www.python.org}), and Python software developed by HERA collaboration members.
Here we provide a list of these packages and their version:
\texttt{aipy [v2.1.12]} (\url{https://github.com/HERA-Team/aipy}), \texttt{healvis [v1.0.0]} \citep[\url{https://github.com/RadioAstronomySoftwareGroup/healvis};][]{Lanman2019b}, \texttt{hera\_cal [v2.0]} (\url{https://github.com/HERA-Team/hera_cal}), \texttt{hera\_sim [v0.0.1]} (\url{https://github.com/HERA-Team/hera_sim}), \texttt{pyuvdata [v1.3.6]} \citep[\url{https://github.com/RadioAstronomySoftwareGroup/pyuvdata};][]{Hazelton2017}, and \texttt{uvtools [v0.1.0]} (\url{https://github.com/HERA-Team/uvtools}).
These packages in turn rely heavily on other publicly available software packages, including:
\texttt{astropy [v2.0.14]} \citep[\url{https://astropy.org};][]{Astropy2013}, \texttt{healpy [v1.12.9]} (\url{https://github.com/healpy/healpy}), \texttt{h5py [v2.8.0]} (\url{https://www.h5py.org/}), \texttt{matplotlib [v2.2.4]} (\url{https://matplotlib.org}), \texttt{numpy [v1.16.2]} (\url{https://www.numpy.org}), \texttt{scipy [v1.2.1]} (\url{https://scipy.org}), and \texttt{scikit-learn [v0.19.2]} (\url{https://scikit-learn.org}).

%% Bibliography %%
\bibliographystyle{apj}
\bibliography{abscal_paper}

\end{document}